\documentclass[sigconf]{acmart}

\setcopyright{acmlicensed}
\copyrightyear{2026}
\acmYear{2026}

\acmConference[ICSE 2026]{48th International Conference on Software Engineering}{April 2026}{Rio De Janeiro, Brazil}

\usepackage{pgfplots}
\pgfplotsset{compat=1.18}
\usepackage{pgfplotstable}
\usetikzlibrary{patterns}
\usepgfplotslibrary{statistics}
\usepackage{tikz}
\usepackage{algorithm}
\usepackage[noend]{algpseudocode}

\tikzset{
	source/.style={rectangle, draw, fill=blue!20, text width=2.5cm, align=center, minimum height=1cm},
	sink/.style={rectangle, draw, fill=green!20, text width=2.5cm, align=center, minimum height=1cm},
	link/.style={thick, draw=black, -{Latex[round]}, opacity=0.7}
}

\usepackage{pgf-pie}
\usepackage{pifont}
\usepackage{rotating}
\usepackage{tikz}
\usetikzlibrary{positioning}
\usepackage{listings}
\usepackage{booktabs}
\usepackage{adjustbox}
\usepackage{multirow}
\usepackage{graphicx}
\usepackage{subcaption}
\usepackage{amsmath}
\usepackage{enumitem}
\usepackage{xcolor}
\usepackage{xspace}
\usepackage{balance}
\usepackage{enumerate}
\usepackage{nameref}

\definecolor{bluebar}{HTML}{376795}
\definecolor{darkgreen}{rgb}{0.0, 0.5, 0.0}
\definecolor{todocolor}{HTML}{FDDA0D}

\usepackage{tcolorbox}
\newcounter{findingCounter}
\usepackage{tikz}

\definecolor{promptbgcolor}{gray}{0.9}
\lstdefinestyle{promptstyle} {
	basicstyle=\ttfamily\small,
	moredelim=**[is][\color{teal}]{@@}{@@} 
}

\newcommand{\review}[1]{\textcolor{black}{#1}}

\newcommand{\approach}{\mbox{\textsc{E-Test}}\xspace}
\newcommand{\approachR}{\mbox{\textsc{E-Test$_{RAG}$}}\xspace}

\newcommand{\tested}{\emph{already-tested}\xspace}
\newcommand{\untested}{\emph{need-test}\xspace}

\newcommand{\errorProne}{\emph{error-prone}\xspace}

\newcommand{\Tested}{\emph{Already-tested}\xspace}

\newcommand{\Untested}{\emph{Need-test}\xspace}
\newcommand{\ErrorProne}{\emph{Error-prone}\xspace}

\newcommand{\gitScenarios}{\textsc{150}\xspace}
\newcommand{\DefectJScenarios}{\textsc{1,825}\xspace}
\newcommand{\allScenarios}{\textsc{1,975}\xspace}
\newcommand{\gitBugs}{\textsc{50}\xspace}
\newcommand{\defectBugs}{\textsc{669}\xspace}
\newcommand{\allBugs}{\textsc{719}\xspace}

\newcommand{\numAllSimilarScenarios}{\textsc{537}\xspace}

\newcommand{\precision}{\textsc{0.55}\xspace}
\newcommand{\recall}{\textsc{0.59}\xspace}
\newcommand{\fonescore}{\textsc{0.55}\xspace}

\newcommand{\gpt}{GPT-3.5 Turbo\xspace}
\newcommand{\llamaeight}{Llama3 8B\xspace}

\newcommand{\gptfturbo}{GPT-4 Turbo\xspace}
\newcommand{\gptfo}{GPT4o\xspace}

\newcommand{\sut}{\emph{MUT}\xspace}

\newcommand{\preprocessor}{\textsc{PreProcessor}\xspace}
\newcommand{\classifier}{\textsc{Analyzer}\xspace}
\newcommand{\postprocessor}{\textsc{PostProcessor}\xspace}

\newcommand{\promptBuilder}{\textit{Builder}\xspace}
\newcommand{\promptTemplate}{\textit{Template}\xspace}
\newcommand{\selector}{\textit{Selector}\xspace}

\newcommand{\voter}{\textit{Voter}\xspace}

\newcommand{\mut}{\emph{MUT}\xspace}
\newcommand{\mutsuite}{\emph{MUT \textsc{Tests}}\xspace}
\newcommand{\mutscenario}{\emph{MUT \textsc{Input}}\xspace}

\newenvironment{summary}[1]{
	\begin{tcolorbox}[colback=cyan!10!white, 
		colframe=cyan!10!white, 
		arc=0mm,grow to left by=0mm,left=0mm,grow to right by=0mm,left=0mm,right=0mm,top=0mm,bottom=0mm]
		
	}
	{
	\end{tcolorbox}
}

\usepackage{hyperref}

\usepackage{booktabs}
\usepackage{graphicx}

\setlist{nolistsep,leftmargin=5ex}

\begin{document}

\title{\approach: E’er-Improving Test Suites}

\author{Ketai Qiu}
\orcid{0009-0002-9750-2762}
\affiliation{
\institution{Università della Svizzera Italiana (USI)}
\city{Lugano}
\country{Switzerland}}
\email{ketai.qiu@usi.ch}

\author{Luca Di Grazia}
\orcid{0000-0002-5306-8645}
 \affiliation{
 	\institution{Università della Svizzera Italiana (USI)}
 	\city{Lugano}
 	\country{Switzerland}}
 \email{work@lucadigrazia.com}

 \author{Leonardo Mariani}
 \orcid{0000-0001-9527-7042}
 \affiliation{
 	\institution{Università Milano-Bicocca}
 	\city{Milan}
 	\country{Italy}}
 \email{leonardo.mariani@unimib.it}

 \author{Mauro Pezzè}
 \orcid{0000-0001-5193-7379}
\affiliation{
\institution{Università della Svizzera Italiana (USI)}
\city{Lugano}
\country{Switzerland};
\institution{Università Milano-Bicocca}
\city{Milan}
\country{Italy};
\institution{Constructor Institute of Technology (CIT)}
\city{Schaffausen}
\country{Switzerland}}

 \email{mauro.pezze@usi.ch}


\begin{abstract}
Test suites are inherently imperfect, and testers can always enrich a suite with new test cases that improve its quality and, consequently, the reliability of the target software system. However, finding test cases that explore execution scenarios beyond the scope of an existing suite can be extremely challenging and labor-intensive, particularly when managing large test suites over extended periods.

In this paper, we propose \approach, an approach that \review{reduces the gap between the execution space explored with a test suite and the executions experienced after testing by augmenting the test suite with test cases that explore execution scenarios that emerge in production. \approach} 
(i) identifies executions that have not yet been tested from large sets of scenarios, such as those monitored during intensive production usage, and (ii) generates new test cases that enhance the test suite. \approach leverages Large Language Models (LLMs) to pinpoint scenarios that the current test suite does not adequately cover, and augments the suite with test cases that execute these scenarios.

Our evaluation on a dataset of \allScenarios scenarios, collected from highly-starred open-source Java projects already in production and Defects4J, demonstrates that \approach retrieves not-yet-tested execution scenarios significantly better than state-of-the-art approaches. While existing regression testing and field testing approaches for this task achieve a maximum F1-score of 0.34, and vanilla LLMs achieve a maximum F1-score of 0.39, \approach reaches \review{0.55}.

These results highlight the impact of \approach in enhancing test suites by effectively targeting not-yet-tested execution scenarios and reducing manual effort required for maintaining test suites. 

\end{abstract}

\maketitle

\section{Introduction}

Test suites are extremely important to guarantee the quality of software systems~\cite{Horgan:Quality:Computer:1994, Zhou:MetamorphicTesting:TSE:2016}.
Test suites are inherently imperfect, that is, they never reveal all bugs in the software system, and can always be improved, no matter how well maintained they are~\cite{
	Kochhar:TestSuiteEffectiveness:SANER:2015, 
	Hindle:JavaTestEvolution:MSR:2023}.
Even mature software systems undergo continuous testing~\cite{
	Memon2017, 
	Zaidman:EngineerTestCases:TSE:2022, 
	Alshahwan:AutomatedUnitTestImprovement:FSE:2024}.
An effective quality process of a long-lasting software product requires test suites that software engineers maintain and evolve throughout the software lifetime~\cite{
	Harrold:SizeOfTestSuite:TOSEM:1993, 
	Pinto:TestSuiteEvolution:FSE:2012, 
	pham2022, 
	winkler2022we}.

So far the research about long-lasting testing has focused mostly on test augmentation, regression and field testing~\cite{Elsner2021, Lei2024, Cheng:TestCasePrioritization:ISSTA:2024}. Test augmentation and regression approaches remove obsolete test cases, select, prioritize and filter test cases to control the size of the suite~\cite{yoo2012regression, Ma:TestSelection:TOSEM:2021, Pan:TestCaseSelection:EMSE:2022, greca2023, Liu2023, Huang:TestCaseSelection:TOSEM:2024}, and augment the suite with test cases that exercise the modified code~\cite{Santelices:TestSuiteAugmentation:ASE:2008}.
Field testing executes software systems in the field to reveal failures that emerge only in production~\cite{gazzola2017exploratory, Caldas:FieldBasedTesting:TSE:2024}, usually executing test cases that instantiate templates in the production environment~\cite{bertolino2021}.

In this paper we tackle the problem of long-lasting testing from a different and more ambitious perspective than regression and field testing. We propose \emph{e'er-improving test suites}\footnote{\emph{E’er} is an ancient English term for \emph{Ever}.}, test suites that automatically improve with data that become available at any time during the software lifecycle (a major source being data from monitoring the execution of the software system in production) to widen the executions that the test suite exercises, thus increasing the coverage of execution scenarios.   
We start from the observation that the execution in production is by far the largest and most complete set of actual execution scenarios of a software system~\cite{Brunetto:AutomaticTestCaseGeneration:JSS:2021}. We observe that the execution scenarios from production offer a unique opportunity to \emph{explore not-yet-tested scenarios}.
Unfortunately, simply augmenting the test suite with all the scenarios observed in the software lifetime blows up the suite that quickly becomes useless~\cite{fraser2012whole}.
We argue that it is indeed possible to continuously improve a test suite, by sifting huge sets of scenarios, like the ones that we obtain from monitoring the execution in production, to identify \untested and \errorProne scenarios, and generate test cases from them.  

We propose \approach, an approach that classifies execution scenarios as \tested, \untested and \errorProne with respect to a test suite, and enriches the test suite with new test cases that exercise \untested and \errorProne scenarios.
\review{An execution scenario for a Software Under Test (SUT) contains the input and the state for the SUT.}
The scenario is \tested with respect to a test suite,  if the input is \emph{``equivalent''} to a test case in the suite.  It is  \untested, if there is \emph{``no equivalent test''} case in the suite and it succeeds, \errorProne if the method fails.

\review{Classifying scenarios by executing all the test cases is impractical for industrial-scale software and test suites. \approach efficiently classifies scenarios without executing the test cases.}
Our assumption is that Large Language Models (LLMs) can effectively classify scenarios without executing all test cases thanks to the huge corpus of public bug reports, issues, code, and tests that belong to the training data of the LLM. However, the use of vanilla LLMs does not produce the expected results, as we report in Section~\ref{sec:evaluation}. 
\approach fine tunes an LLM, defines a set of prompts, queries the LLM with Retrieval-Augmented Generation (RAG) and merges the results of querying the LLM, to overcome the limitations and improve the poor effectiveness of a vanilla LLM.

\approach effectively classifies the execution scenarios, by investigating four key characteristics that characterize the effectiveness of a scenario with respect to a test suite: the ability to improve \emph{coverage}, the \emph{uniqueness of the execution} of the  scenario, the \emph{correctness} of the output, and the likelihood of the scenario to reveal a \emph{bug}.
We carefully engineered a prompt template that assesses an input execution scenario, by asking five questions about the four characteristics, and we defined a strategy that combines the answers to the questions, to effectively classify the scenario as \tested, \untested or \errorProne.

We created a dataset of \allScenarios scenarios from both popular Java projects available on GitHub and extending Defects4J~\cite{Just2014}. We used the dataset to experimentally evaluate both the effectiveness of \approach to classify execution scenarios and the sensitivity of \approach to different configurations\footnote{\approach is available on a replication package at \href{https://github.com/ketaiq/E-Test-package}{https://github.com/ketaiq/E-Test-package} to facilitate independent replicas of the experiments.}.
The experimental results indicate that \approach achieves \review{\precision} precision, \review{\recall} recall, \review{\fonescore} F1-score, on average.
We compared \approach to state-of-the-art approaches for both regression testing and field testing, \textit{FAST++}~\cite{Cruciani:ScalableApproachesForTestSuiteReduction:ICSE:2019} and \textit{field-ready testing}~\cite{gazzola2022}, respectively.
The comparative evaluation indicates that \approach outperforms both \textit{FAST++} and \textit{field-ready testing} in terms of F1-score, with a relative increase of \review{61.8\%} with respect to \textit{FAST++}  and \review{83.3\%} with respect to \textit{field-ready testing}. 
We automatically generated test cases for all Defects4J scenarios that \approach classifies as \errorProne, to check for the effectiveness of \approach to generate error-revealing test cases.  The generated test cases detect 83.2\% of the failures documented for the considered scenarios in Defects4J.

This paper contributes to state of the art in software testing by:

\begin{itemize}
	\item spotlighting a novel viewpoint of \emph{e'er-improving test suites}: improving test suites by identifying execution scenarios that can enrich the suite,
	\item defining \approach, an approach that leverages LLMs with prompts, RAG, fine-tuning, and merging of the results of multiple queries, to classify scenarios as \tested, \untested, or \errorProne, and thus to identify new candidate test cases,
	\item creating a dataset of \allScenarios execution scenarios that we harvested from both Defects4J and four popular Java projects available on GitHub, to evaluate \approach,
	\item discussing the results of a thorough experimental evaluation of \approach across different Java projects, for comparatively evaluating \approach with respect to vanilla LLMs and state-of-the-art regression and field testing approaches,
\item presenting a test case generator from execution scenarios, and evaluating the ability of \approach to reveal failures. 
\end{itemize}

\smallskip

The paper is organized as follows. \review{Section~\ref{sec:approach} defines \approach and introduces a running example.}
Section~\ref{sec:evaluation} discusses the experimental results. 
Section~\ref{sec:related} overviews related work. Section~\ref{sec:conclusion} summarizes the main results and spotlights the open research directions. 

\section{\approach}
\label{sec:approach}

\emph{E’er-improving test suites} are long-lasting test suites that automatically grow with new test cases that
incrementally explore new partitions of the execution space, that is, exercise behaviors that the original test suite misses.  
\approach automatically identifies test cases that exercise unexplored behaviors from sets of execution scenarios, by classifying scenarios as \tested, \untested or \errorProne. 
\review{When fed with execution scenarios from production, \approach efficiently identifies behaviors that are observed in production and not well-tested yet. Thus, it reduces the gap between the tested and production spaces, that is, the executions that the test suite exercises and the executions in production.}

\review{\approach works with a SUT, a test suite for the SUT, and the input for the SUT, and indicates whether the input exercises a new behavior and thus is a good candidate to generate a test case that improves the test suite. The definition of \approach does not restrict the application of the approach to a specific testing level. In principle, \approach can be instantiated for unit, integration and system testing. 
In this paper, we present \approach for augmenting unit test suites, to benefit from large and widely used benchmarks for fine-tuning the core LLM component of \approach.  Adapting \approach to integration and system testing requires suitable benchmarks for fine-tuning.
} 

An execution scenario for a method is a tuple $\langle In, M, T\rangle$, where $In$ is an input for a method $M$, $M$ is a method with its \emph{context data}, and $T$ is a test suite for $M$. The context data for a method $M$ are the \emph{operations executed to build the objects} that method $M$ uses as parameters and accesses as external variables, if any.
\review{
For example, $\mathcal{S}$ is an execution scenario for the method \texttt{asNode} in \texttt{Spring Boot}\footnote{\url{https://github.com/spring-projects/spring-boot}}:
\begin{align*}
	\mathcal{S} = \langle & \texttt{"[::1]:6379"},\ \texttt{asNode},\ \mathcal{T} \rangle \\
	\mathcal{T} = \{ &\texttt{asNode("127.0.0.1:1111")},\ \texttt{asNode("127.0.0.2:2222")}, \\
	& \texttt{asNode("127.0.0.3:3333")} \}
\end{align*}
where \texttt{"[::1]:6379"} is the input and $\mathcal{T}$ is the test suite for the method \texttt{asNode}.
We use this scenario that led to a failure in production (GitHub issue \#39819\footnote{\url{https://github.com/spring-projects/spring-boot/pull/39819}}, diagnosed on March 1, 2024) as a running example in this paper.
}

An execution scenario $\langle In, M, T\rangle$ is \tested, if $In$ is “\emph{equivalent}’’ to a test in $T$. Two test cases are equivalent if they belong to the same partition, from the partition testing viewpoint~\cite{Richardson:partion:ICSE:1981}. We recall that partition testing \emph{``divides the infinite set of possible test cases into a finite set of classes (partitions), with the purpose of drawing one or more test cases from each class''} and that each partition groups test cases with uniform behavior, that is, test cases that exercise the same portion of the program, according to some test case selection criterion~\cite{pezze2008}. The identification of partitions depends on the test case selection criterion. In our experiments, we refer to a combination of structural and functional criteria that lead to the definition of the set of heterogeneous queries.

An execution scenario $\langle In, M, T\rangle$ is either \untested or \errorProne, if $In$ is \emph{``not equivalent''} to a test in $T$ for $M$, and $M$ either succeeds or fails when executed with $In$, respectively. \Tested scenarios increase the size of the suite without enriching it. \Untested scenarios enrich the test suite by sampling not-yet-tested partitions, thus augmenting the quality of the test suite. \ErrorProne scenarios reveal bugs to be pruned from the code.

\review{
The example scenario $\mathcal{S}$ executes the method \texttt{asNode} with an input IPv6 address (\texttt{"[::1]:6379"}), an IP format that is not exercised with the test cases in $\mathcal{T}$ that use the IPv4 format (\texttt{x.x.x.x:xxxx}). The method \texttt{asNode} fails in this scenario, due to a parsing error, as documented in the GitHub issue. Thus, the scenario $\mathcal{S}$ is \errorProne, since it exercises a scenario different from the scenarios that the suite exercises, and causes the method to fail.
The scenario $\langle \texttt{"127.0.0.1:5678"},\ \texttt{asNode},\ \mathcal{T} \rangle$ is \tested with respect to the test suite $\mathcal{T}$, since it exercises executions that the tests in $\mathcal{T}$ already exercised, and the method \texttt{asNode} does not fail with the input \texttt{"127.0.0.1:5678"}.
We observe that the scenario $\mathcal{S}$ becomes \untested for the method \texttt{asNode} after fixing the bug, since it exercises an execution that the tests in $\mathcal{T}$ do not exercise, and the fixed method does not fail.
}

\begin{figure*}[t]
	\centering
	\includegraphics[width=.95\linewidth]{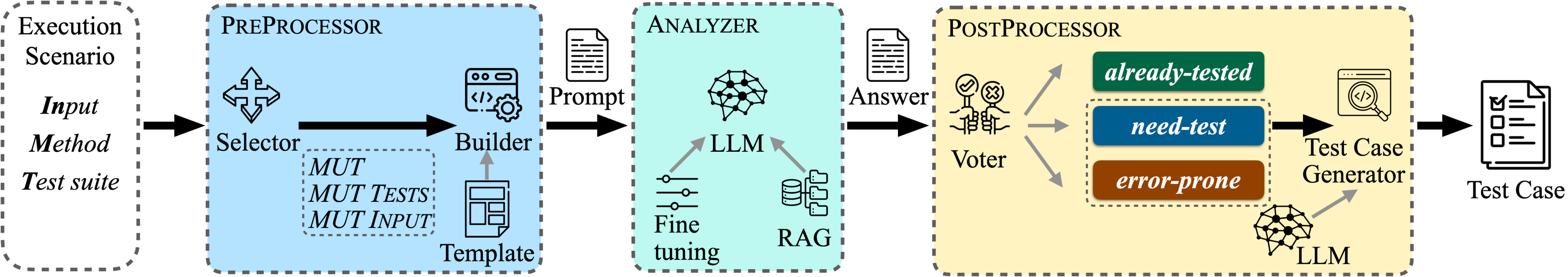}
	\caption{Overview of \approach}
	\label{fig:overview}
\end{figure*}

\review{Figure~\ref{fig:overview} illustrates the three phases of \approach: \preprocessor, \classifier, and \postprocessor. The \preprocessor builds the prompts for the \classifier from the execution scenario. The \classifier queries an LLM to answer the prompt. The \postprocessor classifies the input scenario as \tested, \untested, or \errorProne, and generates test cases for both \untested and \errorProne scenarios.
}

\begin{figure}[t]
	\centering
	\includegraphics[width=.95\linewidth]{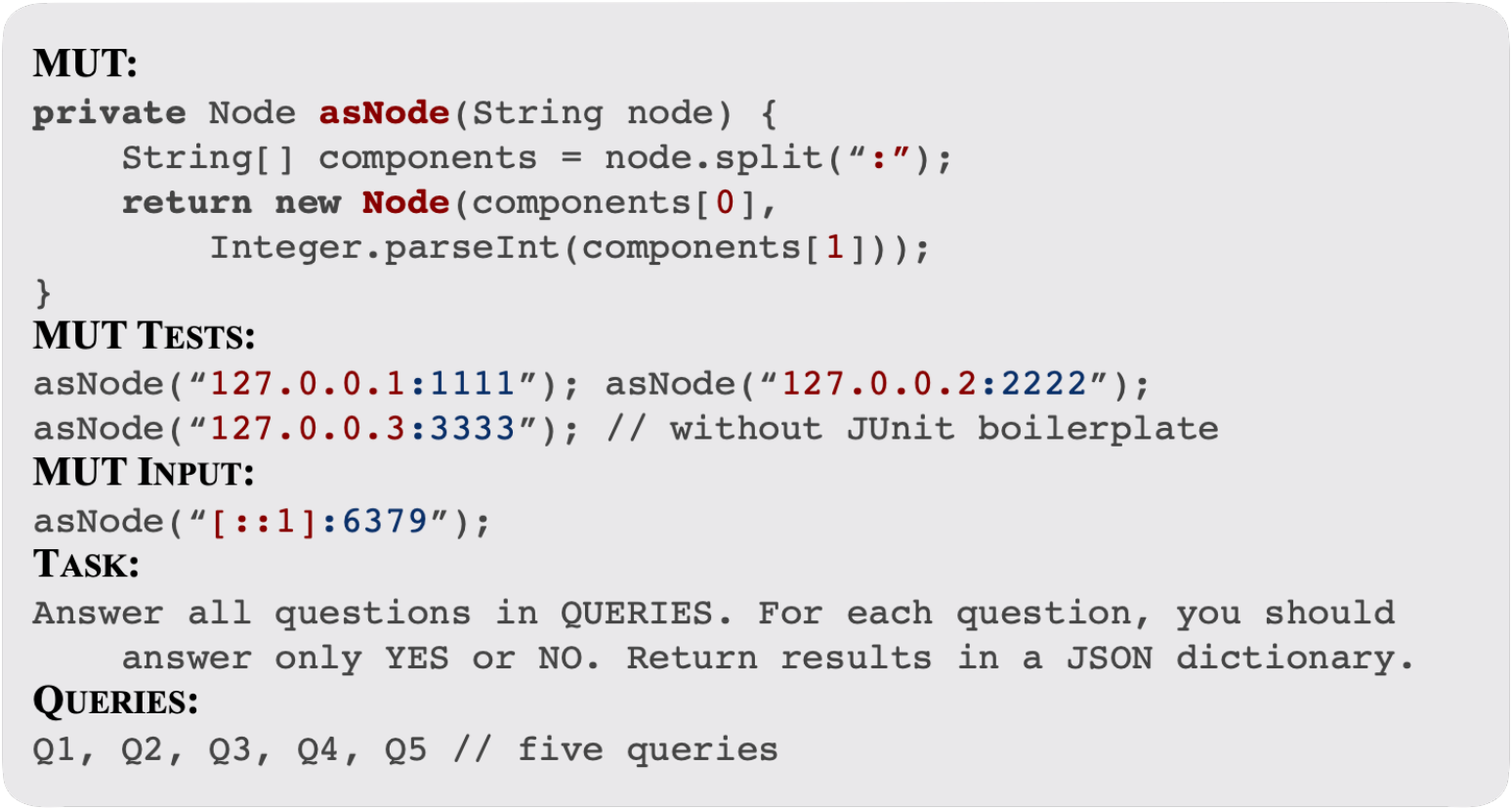}
	\caption{\review{A structured prompt for the example scenario $\mathcal{S}$ instantiated by \approach}}
	\label{fig:example_prompt}
\end{figure}

\subsection{\preprocessor}
\label{sec:preprocessor}

The \preprocessor generates a structured prompt  by combining a \textit{Selector} and a \promptBuilder. 
The \textit{Selector} collects  the \mutscenario, the \mut, and the \mutsuite. The \promptBuilder builds the prompt by instantiating a template on the three items and the queries.

\subsubsection*{Selector}
The \textit{Selector} mines three essential items from the input execution scenario: the \emph{\textbf{M}ethod \textbf{U}nder \textbf{T}est} \mut (less than 3K tokens, to comply with the LLMs we currently use in our experiments), the \mutsuite (less than 4K tokens), and the \mutscenario (less than 1K tokens).  Selecting the items with upper bounds is important to both match the maximum context length of the LLM and optimize the cost of the query that depends on the number of tokens~\cite{wang2024}. 
\review{Figure~\ref{fig:example_prompt} shows the \mut, \mutsuite and \mutscenario that the \textit{Selector} produces for scenario $\mathcal{S}$.
}

The \mut is the code snippet relevant for understanding the focal method $M$: the signature and body of $M$, and the strongly related methods that $M$ directly invokes and that belong to the same Java class of $M$.
\selector retrieves such code snippet with JavaParser~\cite{javaparser} to statically collect Java code within the method call dependency graph.
\emph{Selector} selects the \mutsuite as the set of test cases that directly or indirectly (within 3 nested calls) execute $M$ from the test suite of the class of $M$.
We truncate the characters of the test suite to 4K tokens to fit LLM's context limit.
\emph{Selector} collects the \mutscenario, that is, the parameters and the surrounding context code of the call to $M$, from the Java bytecode with ASM\footnote{\url{https://asm.ow2.io/}}, a popular framework for instrumenting bytecode~\cite{Li:DJXPerf:CGO:2023}. 
The Java instrumentation is not the only possible strategy to collect \mutscenario, but it is also possible to integrate with different carving techniques~\cite{Gambi2023}. 
\emph{Selector} compares the signature of $M$ with the instrumented methods to retrieve the direct method call, encoded as a string. 
It extends the method call string with the surrounding context, that is, the code within the ten lines both before and after the method call.

The information about the context of the method is extremely useful for the LLM to understand the invocation.
For example, \emph{Selector} builds the \mutscenario of method \texttt{solve} in Defects4J, by augmenting the method call \texttt{solver.solve(f, Math.PI, 4)} with the statements \texttt{Univariate\-Real\-Function f = new SinFunction()} and \texttt{Univariate\-Real\-Solver solver = new BrentSolver()}, the code snippets that \emph{Selector} extracts from the surrounding context code of \texttt{solve}, and that build the actual parameters of the call.

\subsubsection*{Builder}

The \promptBuilder instantiates the five sections of the \promptTemplate
\review{by augmenting \mut, \mutsuite and \mutscenario that the \textit{Selector} produces with \textit{\textsc{Task}} and \textit{\textsc{Queries}}.
The five sections in Figure~\ref{fig:example_prompt} shows the \promptTemplate that the \promptBuilder produces for scenario $\mathcal{S}$:
}

\begin{description}

	\item[MUT:] The source code of the method \texttt{asNode(String node)}, which is the method under test $M$.

	\item[MUT \textsc{Tests}:] The test suite of \texttt{asNode}, that \textit{Selector} extracts from the suite of Class \texttt{Properties\-Redis\-Connection\-Details}, that includes \texttt{asNode}. 

	\item[MUT \textsc{Input}:] The method call of \texttt{asNode (String node)} with the input \texttt{“[::1]:6379”} that we ask \approach to classify.  The simple call in the example does not require context information. 

	\item[\textsc{Task}:]
	The instruction about answering five queries for the LLM: 
	Given \mut, \mutsuite and \mutscenario, we ask LLM to answer the questions in \textit{Queries}. For each question, we ask LLM for a binary \textsc{Yes} or \textsc{No} answer in JSON format.
	 
	\item[\textsc{Queries}:] The five queries that we ask LLM to answer to assess \mutscenario and that we discuss in the next section.
\end{description}

\subsubsection*{Queries}
\label{subsubsec:queries}
We assume that LLMs can effectively infer the characteristics of scenarios, thanks to the huge corpus of public bug reports, issues, code, and tests that belong to the training data of the LLM.  
We formulate the queries about the input scenario, by considering the aspects of a scenario that are likely related to the execution space that the suite tests: the similarities of the scenario with respect to the test cases in the suite, the differences of the execution of the scenario with respect to the behaviors that the suite exercises, the likelihood of a correct result or failure.

We formulate five queries about different although partially overlapping characteristics.

\begin{description}
	\item[\textbf{Q1:}] \textbf{Is \mutscenario a similar scenario compared with \mutsuite?} This query concerns the  \emph{logical similarity} of  the \mutscenario with respect to the scenarios that the suite tests. Thus it aims to distinguish scenarios that are already tested  from  scenarios that deserve to be further tested.

	\item[\textbf{Q2:}] \textbf{Does \mutscenario cover more lines or branches than \mutsuite?} This query complements Q1 by highlighting the differences in code coverage.
	The extent to which the \mutscenario exercises new code elements with respect to the \mutsuite heuristically identifies areas of the code that the suite does not test yet.
	
	\item[\textbf{Q3:}]  \textbf{Will \mut work differently when executed under \mutscenario?} This query looks for \emph{anomalous behaviors} that the \mutscenario reveals.  Anomalies in the execution are key elements that deserve particular attention when testing a software.	

	\item[\textbf{Q4:}]  \textbf{Does \mut still produce correct results when executed under \mutscenario?} This query estimates the correctness of the \emph{result} of executing the \mut with the \mutscenario, and spotlights scenarios that deserve further testing. 
	
	\item[\textbf{Q5:}]  \textbf{Will \mutscenario reveal any bug in \mut?} This query reinforces Q4, by stressing the possibility of a failure, by asking LLMs to focus on potential bugs in the code.
\end{description}

We formulate the queries with heterogeneous answers with respect to the classification task (the relevant answer for characterizing the senario as relevant for testing may be positive or negative, depending on the question), to reduce the risk of LLMs following a fixed answering pattern, with undesirable biases in the answers.   By mixing the answers  that identify each category, we encourage LLMs to evaluate each query on its own merits. As a result, the expected answers for classifying a specific scenario are not all \textsc{Yes} or all \textsc{No}, but a combination of \textsc{Yes} or \textsc{No}. 

We define questions that LLM can properly understand and answer,  by looking for terms that might have been extensively used in the training set of the LLM in many different contexts, and thus with a meaning that LLM can easily interpret. To this end, we identified the most suited terms for the queries about software bugs from Stack Overflow\footnote{\url{https://stackoverflow.com/}}, by assuming a key role of Stack Overflow in the training of popular LLMs.  
We sorted the queries according to the upvote.  We selected all terms that occur in the top 10\% queries, filtered out the stop words, sorted the terms by frequency, and selected the top terms related to the key terms in the queries. We obtained \emph{similar}, \emph{cover}, \emph{line}, \emph{branch}, \emph{differently}, \emph{correct}, and \emph{reveal}. We  use these terms to build the queries.

\subsection{\classifier}
\label{sec:classifier}

The \classifier queries an LLM to answer the \review{five queries of the prompt. The answers to the five queries for the prompt corresponding to the scenario $\mathcal{S}$ in JSON format are $\{$Q1: NO, Q2: YES, Q3: YES, Q4: NO, Q5: YES$\}$. The input is not similar to any test case in the test suite (Q1: NO), it likely increases the coverage of the test suite (Q2: YES), it likely works differently from any test case in the test suite (Q3: YES), it likely  does not produce a correct result (Q4: NO), and it likely reveals a bug in \texttt{asNode} (Q5: YES).
}
\approach implements LLMs  with  fine-tuning~\cite{Weyssow:FineTuningCodeGenerationLargeLanguageModels:TOSEM:2025}, and Retrieval-Augmented Generation (RAG) \cite{Lewis:RetrievalAugmentedGeneration:NIPS:2020}, as illustrated in Figure~\ref{fig:overview}. 
Here we present the dataset used to validate \approach and detail the techniques implemented within it.

\subsubsection*{Dataset}
\label{sec:dataset}

\begin{table*}[t]
	\centering
	\footnotesize
	\review{
	\caption{\review{The GitHub dataset used in the experiments}}
	\label{tab:project-stats}
	\resizebox{\textwidth}{!}{%
	\begin{tabular}{l l r r r r r r}
		\toprule
		\textbf{Project} & \textbf{Version} & \textbf{LOC} & \textbf{\#Classes} &
		\textbf{\#Test Suites} & \textbf{\#Test Cases} &
		\textbf{Avg.\ Test Cases / Suite} & \textbf{Avg.\ LOC / Suite} \\
		\midrule
		Spring Boot             & 3.1.12 & 683\,988  & 6\,901 & 2\,261 & 13\,328 & 5 & 135 \\
		Apache ShardingSphere   & 5.5.0  & 1\,519\,076 & 7\,635 & 1\,382 & 5\,867 & 4 &  92 \\
		Apache Dolphinscheduler & 3.2.1  & 257\,915  & 2\,336 &   469 & 1\,988 & 4 & 143 \\
		Micrometer              & 1.11.12 & 166\,565  & 1\,231 &   385 & 1\,684 & 4 & 191 \\
		\bottomrule
	\end{tabular}
	}}
\end{table*}

Our dataset consists of \allScenarios execution scenarios that we obtained by augmenting an initial set of \DefectJScenarios scenarios from Defects4J with \gitScenarios scenarios that we mined from GitHub. The generation of the dataset requires about 180 person-hours, to build a dataset reusable in further studies.

Defects4J is a popular benchmark with all the information required to understand, test and replicate software bugs that have been reported from the field, after software has been publicly released~\cite{Just2014}.  
We augmented the scenarios that we obtained from Defects4J with a set of scenarios that we collected by sampling GitHub, to increase the diversity of the dataset with issues about additional libraries and projects.

We mined the closed GitHub issues from highly-starred open-source Java projects (Spring Boot, Apache ShardingSphere, Apache Dolphinscheduler and Micrometer~\footnote{
	\href{https://github.com/spring-projects/spring-boot}{Spring Boot}, 
	\href{https://github.com/apache/shardingsphere}{ShardingSphere}, 
	\href{https://github.com/apache/dolphinscheduler}{Dolphinscheduler}, and 
	\href{https://github.com/micrometer-metrics/micrometer}{Micrometer} on GitHub}). \review{Table~\ref{tab:project-stats} shows the statistics of the GitHub dataset that we used in the experiments (version, LOC, number of classes, number of test suites, number of test cases, and averages).}
 We filtered the results by label (\emph{bug} or \emph{defect}), creation date (\emph{after 1 January 2020}), and number of comments (\emph{at least 5}). We carefully selected \gitBugs non-trivial and reproducible bugs that were triggered due to not-yet-tested inputs. 

We created a set of \allBugs \errorProne scenarios $\langle In, M, T \rangle$ by reproducing  \defectBugs Defects4J bugs\footnote{We discarded 166 bugs where \mut or \mutscenario exceeded the context limit.} using the available tests and the \gitBugs bugs from GitHub.
We built the \untested scenarios from the \errorProne scenarios by (i) replacing the buggy method $M$ with the repaired method,  (ii) reproducing the same execution scenario, and (iii) checking that the test passes.  The scenarios that we obtain in this way are missing execution scenarios by construction, since the original fault escapes the testing. 
The scenarios are  not \errorProne since we build them with the repaired method.

We completed the dataset with \numAllSimilarScenarios \tested scenarios that we obtained by (1) substituting the input $In$ with a new test input that (i) we generated with Evosuite~\cite{fraser2011}, (ii) is not in the test suite $T$, (iii) succeeds, and (iv) does not increase the branch coverage of $T$, and (2) keeping the buggy $M$.  The scenarios that we obtain in this way do not contribute to improving the test suite.

The final dataset contains \allBugs \errorProne scenarios, \allBugs \untested scenarios, and \numAllSimilarScenarios \tested scenarios.  The dataset is available in the replication package~\footnote{\url{https://github.com/ketaiq/E-Test-package}}.

\subsubsection*{Fine-tuning}

Vanilla LLMs often do not perform well for specific tasks, due to their general-purpose nature and training~\cite{Wang:SurveyFewshotLearning:ACMCS:2020}. \citet{Chow2024} observe that pretrained LLMs do not usually work best on code-related tasks, and show that the performance of LLMs improves when the LLMs are fine-tuned.  The requirements of software testing demand a level of precision and contextual understanding that general LLMs may not provide~\cite{wang2024}. Fine-tuning  the LLM can improve its capability to accurately classify execution scenarios.

We fine-tuned the LLM with samples from all the three types of scenarios: \tested, \untested, and \errorProne. Each sample consists of a tuple of $\langle \text{\emph{Prompt}, \emph{Responses}} \rangle$, where \emph{Prompt} is the generated prompt for a scenario \(\langle In, M, T \rangle\), and \emph{Responses} is the ground-truth answers to the five queries in JSON format.

We fine-tuned only \gpt, since it is one of the best vanilla LLMs for predicting not-yet-tested scenarios.
We fined-tuned \gpt with samples from all 17 projects of Defects4J. We split the Defects4J part of our dataset into \emph{fine-tuning} and \emph{validation} sets with a 5:95 ratio, and with balanced sets of scenarios of the three types within the \emph{fine-tuning} set.
We fine-tuned the model with batch size 1, learning rate multiplier~2.0 and 3 epochs using the balanced fine-tuning dataset.

\subsubsection*{RAG Retrieval-Augmented Generation}
\label{subsubsec:rag}

The size limitations of the prompts of vanilla LLMs do not allow to feed the whole set of information in the presence of large  code and test suites.  We implemented RAG to query the LLM with the whole code base, to overcome the size limitations of the prompts. 
We built RAG indices for both the source code and the tests with LlamaIndex\footnote{\url{https://www.llamaindex.ai/}}, to feed LLMs with complete information about the software project.
RAG extends  the \mut, \mutsuite, and \mutscenario components of the prompt with code embedding that is strongly related to the prompt in terms of cosine similarity.  
We embedded code vectors for RAG querying with the OpenAI embedding model \texttt{text-embedding-ada-002}.

\subsubsection*{Few-shot Learning}

Vanilla LLMs may not be very precise when fed with plain prompts only.  We augmented the LLM with few-shot learning~\cite{Wang:SurveyFewshotLearning:ACMCS:2020}, by adding fixed \tested, \untested and \errorProne scenarios that we suitably label, to the head of the prompt, aiming to improve the knowledge we feed to the LLM.
We experimented with 3-shot, 6-shot and 9-shot learning, that is, with 1,2 and 3 scenarios per type, respectively. 
The \classifier queries the LLM with the fixed scenarios and the corresponding ground truth answers from the Defects4J dataset.

\subsection{\postprocessor}

The \postprocessor classifies the scenario as  \tested, \untested, or \errorProne, and generates test cases.
The \voter classifies the scenarios based on the highest number of correct answers among the five queries, as shown in Table~\ref{tab:voting}. 
The \voter classifies a scenario as \tested if it is similar to the already-tested scenarios (Q1 = Yes), does not increase the code coverage (Q2 = No), does not work differently from the test cases in the suite (Q3 = No),  produces a correct output (Q4 = Yes), and does not expose a bug (Q5 = No), according to the LLM. 
The \voter classifies a scenario as \untested if it differs from all the already tested scenario (Q1 = No), increases code coverage (Q2 = Yes), does not work differently form the test cases in the suite (Q3 = No), produces a correct output (Q4 = Yes), and does not expose a bug (Q5 = No), according to the LLM. 
The \voter classifies a scenario as \errorProne if it differs from all the already tested scenario (Q1 = No), increases code coverage (Q2 = Yes), works differently form the test cases in the suite (Q3 = Yes),  produces a wrong output (Q4 = No), and exposes a bug (Q5 = Yes), according to the LLM.

The \voter classifies a scenario according to the highest number of answers that match the values in Table~\ref{tab:voting}.
The mismatch of the answer with respect to the expected values in Table~\ref{tab:voting} may depend on the inaccuracy of the prediction in the answers and the fading boundary between executions and imperfect matching, for instance, a \untested scenario may derive from a test that is dissimilar to the existing ones even if it is not likely to cover new code. 
The \voter solves ties with the priority  \errorProne $>$ \untested $>$ \tested, to ensure not to miss scenarios that cannot be clearly classified, thus privileging the robustness over the size of the suite.

\begin{table}[t]
	\centering
	
	\caption{Truth table for each query and scenario}
	
		\begin{tabular}{@{}lccccc@{}}
			\toprule
			\textbf{Scenario} & \textbf{Q1} & \textbf{Q2} & \textbf{Q3} & \textbf{Q4} & \textbf{Q5} \\ \midrule
			\Tested      & \textsc{Yes} & \textsc{No}  & \textsc{No}  & \textsc{Yes} & \textsc{No}  \\
			\Untested    & \textsc{No}  & \textsc{Yes} & \textsc{No}  & \textsc{Yes} & \textsc{No}  \\
			\ErrorProne  & \textsc{No}  & \textsc{Yes} & \textsc{Yes} & \textsc{No}  & \textsc{Yes} \\ \bottomrule
		\end{tabular}
	
	\label{tab:voting}
\end{table}

The \emph{Test Case Generator} generates JUnit test cases for the \sut from the identified not-yet-tested scenarios, by continuing the conversation with the LLM.

\review{For example, the \voter correctly classifies the scenario $\mathcal{S}$  as  \errorProne, since the \classifier answers $\langle$NO, YES, YES, NO, YES$\rangle$.  \approach augments the original test suite $\mathcal{T}$ with a new test case that reveals the bug.
}

\section{Experimental Validation}
\label{sec:evaluation}

\begin{table*}[ht]
	\centering
	\footnotesize
	\caption{Precision (P), Recall (R), F1-score (F1) for different models and configurations}
	\resizebox{\linewidth}{!}{%
	\begin{tabular}{l||ll||rrr||rrr|rrr||r||r}
		\toprule
		&&\multirow{3}{*}{\textbf{Approach}}&&&& \multicolumn{7}{c||}{\textbf{\emph{Not-yet-tested}}} & \textbf{Total} \\ 
		
		&&& \multicolumn{3}{c||}{\textbf{\Tested}} 
		& \multicolumn{3}{c|}{\textbf{\Untested}} 
		& \multicolumn{3}{c||}{\textbf{\ErrorProne}} 
		& \textbf{Avg.}
		& \textbf{Avg.} \\
		
		&&& \textbf{P} & \textbf{R} & \textbf{F1} 
		& \textbf{P} & \textbf{R} & \textbf{F1} 
		& \textbf{P} & \textbf{R} & \textbf{F1} 
		& \textbf{F1} 
		& \textbf{F1} \\ \midrule \midrule
		
		\multirow{21}{*}{\textbf{RQ1}} & \multirow{12}{*}{\textbf{Vanilla}}
& \review{Llama3 1B} & \review{0.29} & \review{0.80} & \review{0.43} & \review{0.39} & \review{0.11} & \review{0.17} & \review{0.40} & \review{0.16} & \review{0.22} & \review{0.19} & \review{0.27} \\
&& \review{Llama3 3B} & \review{0.20} & \review{0.00} & \review{0.00} & \review{0.36} & \review{0.33} & \review{0.34} & \review{0.37} & \review{\textbf{0.68}} & \review{0.48} & \review{0.41} & \review{0.28} \\
		&& Llama3 8B                & 0.26 & 0.19 & 0.22 & 0.36 & 0.51 & 0.42 & 0.36 & 0.27 & 0.31 & 0.37 & 0.32 \\
		&& Llama3 70B               & 0.18 & 0.29 & 0.22 & 0.34 & 0.28 & 0.31 & 0.28 & 0.19 & 0.23 & 0.27 & 0.25 \\

&& \review{Deepseek R1 1.5B} & \review{0.27} & \review{0.49} & \review{0.35} & \review{0.35} & \review{0.19} & \review{0.24} & \review{0.38} & \review{0.33} & \review{0.35} & \review{0.30} & \review{0.31} \\
&& \review{Deepseek R1 7B} & \review{0.18} & \review{0.12} & \review{0.14} & \review{0.36} & \review{0.44} & \review{0.39} & \review{0.36} & \review{0.40} & \review{0.38} & \review{0.38} & \review{0.30} \\
&& \review{Deepseek R1 14B} & \review{0.31} & \review{0.51} & \review{0.37} & \review{0.41} & \review{0.21} & \review{0.27} & \review{0.38} & \review{0.38} & \review{0.38} & \review{0.32} & \review{0.34} \\
&& \review{Deepseek R1 32B} & \review{0.21} & \review{0.22} & \review{0.20} & \review{0.36} & \review{0.40} & \review{0.38} & \review{0.34} & \review{0.30} & \review{0.31} & \review{0.34} & \review{0.30} \\
&& \review{Deepseek R1 70B} & \review{0.24} & \review{0.56} & \review{0.33} & \review{0.35} & \review{0.10} & \review{0.14} & \review{0.34} & \review{0.27} & \review{0.28} & \review{0.21} & \review{0.25} \\      
		&& GPT-3.5 Turbo            & 0.26 & 0.10 & 0.15 & 0.35 & 0.64 & 0.46 & 0.36 & 0.23 & 0.28 & 0.37 & 0.30 \\
		&& GPT-4 Turbo              & 0.17 & 0.20 & 0.18 & 0.34 & 0.56 & 0.42 & 0.31 & 0.07 & 0.11 & 0.26 & 0.24 \\
		&& GPT-4o                   & 0.19 & 0.33 & 0.24 & 0.32 & 0.38 & 0.16 & 0.37 & 0.10 & 0.35 & 0.25 & 0.25 \\
		\cmidrule{2-14}

		& \multirow{3}{*}{\textbf{Few-shot learning}} & GPT-4 Turbo (3-shot)     & 0.19 & 0.24 & 0.21 & 0.35 & 0.52 & 0.42 & 0.32 & 0.09 & 0.14 & 0.28 & 0.26 \\
		&& GPT-4 Turbo (6-shot)     & 0.15 & 0.19 & 0.17 & 0.35 & 0.55 & 0.43 & 0.32 & 0.08 & 0.13 & 0.28 & 0.24 \\
		&& GPT-4 Turbo (9-shot)     & 0.17 & 0.24 & 0.20 & 0.34 & 0.54 & 0.42 & 0.30 & 0.04 & 0.07 & 0.24 & 0.23 \\
		\cmidrule{2-14}
		
		& \multirow{5}{*}{\textbf{RAG}} & Llama3 8B$_{RAG}$       & 0.23 & 0.28 & 0.26 & 0.43 & 0.51 & 0.47 & 0.53 & 0.37 & 0.44 & 0.41 & 0.39 \\
		&& Llama3 70B$_{RAG}$    & 0.22 & 0.51 & 0.30 & 0.34 & 0.22 & 0.27 & \textbf{0.62} & 0.32 & 0.42 & 0.35 & 0.33 \\
		&& GPT-3.5 Turbo$_{RAG}$   & 0.19 & 0.08 & 0.11 & 0.42 & \textbf{0.71} & 0.53 & 0.60 & 0.40 & 0.48 & 0.51 & 0.37 \\
		&& GPT-4 Turbo$_{RAG}$     & 0.15 & 0.34 & 0.21 & 0.31 & 0.33 & 0.32 & 0.55 & 0.11 & 0.19 & 0.26 & 0.24 \\
		&& Deepseek R1 70B$_{RAG}$          & 0.18 & 0.29 & 0.22 & 0.29 & 0.24 & 0.26 & 0.52 & 0.45 & 0.48 & 0.37 & 0.32 \\
		\cmidrule{2-14}

		& \textbf{Baseline} & Random Classifier         & 0.36 & 0.33 & 0.35 & 0.36 & 0.33 & 0.34 & 0.28 & 0.33 & 0.30 & 0.33 & 0.33 \\

		\midrule \midrule
		
		\multirow{2}{*}{\textbf{RQ2}} &\multirow{2}{*}{\textbf{State-of-the-art}} & \citet{Cruciani:ScalableApproachesForTestSuiteReduction:ICSE:2019} 
		& 0.32 & 0.32 & 0.32 & 0.27 & 0.27 & 0.27 & 0.44 & 0.44 & 0.44 & 0.35 & 0.34 \\ 
		&& \citet{gazzola2022}       & 0.50 & \textbf{1.00} & 0.67 & 0.00 & 0.00 & 0.00 & 0.18 & 0.33 & 0.23 & 0.12 & 0.30 \\
		
		\midrule \midrule
		
		& \multirow{2}{*}{\textbf{Fine-tuning}} & \approach                & \review{\textbf{0.67}} & 0.94 & \textbf{0.78} & \review{0.49} & \review{0.26} & \review{0.34} & \review{0.49} & \review{0.58} & \textbf{0.53} & \review{0.43} & \review{\textbf{\fonescore}} \\
		&& \approachR       & 0.47 & 0.41 & 0.44 & \textbf{0.51} & 0.61 & \textbf{0.56} & 0.52 & 0.46 & 0.49 & \textbf{0.53} & 0.50 \\
		\bottomrule
	\end{tabular}
	}
\flushleft
	\label{tab:rq1}
\end{table*}

\subsection*{Research Questions}
We experimentally evaluate \approach to answer the following research questions (RQs):

\begin{description}
	\item[RQ1 Impact of LLMs:] \review{\textbf{What impact does the choice of the LLM have on \approach?}}
	We compare the impact of different pre-trained LLMs on \approach, to measure the contribution of \approach over vanilla LLMs. 
	
	\item[RQ2 Comparative Evaluation:] \textbf{Is \approach better than state-of-the-art approaches?}
	We  compare \approach with state-of-the-art regression testing and field testing approaches, the closest approaches to \approach.  
	
	\item[RQ3 Impact of  Queries:] \review{\textbf{Do the different combinations of queries impact the F1-score?}}
	We evaluate the impact of different combinations of queries (including a single generic query) on the results of \approach, to identify the most effective set of queries. 
	
	\item[RQ4 Efficiency:] 	\textbf{How efficiently does \approach process scenarios?} We evaluate the efficiency of \approach in terms of response time and token consumption of the LLM core, and we compare the different LLMs that we experimented with.

	\item[RQ5 Test Case Generation:] \textbf{Does \approach generate useful test cases?}
	We evaluate the effectiveness of the JUnit test cases that \approach generates.
	
\end{description}

\subsection*{Experimental Setup}
\subsubsection*{Implementation}
We implemented \approach in Python and Java,
with \review{GPT family (3.5Turbo, 4Turbo, 4o), DeepSeek R1 family (1.5B, 7B, 14B, 32B and 70B) and Llama3 family (1B, 3B, 8B, and 70B)}
\footnote{\url{https://openai.com/}, \url{https://www.deepseek.com/} and \url{https://www.llama.com/}}, and we executed the experiments on an Ubuntu 20.04 cluster with four NVIDIA A100 GPUs, each with 40 GB of VRAM. We prompted \gpt, \gptfturbo and \gptfo with the OpenAI APIs, and \review{DeepSeek R1 and Llama3 families} with Ollama APIs~\footnote{\url{https://github.com/ollama/ollama}} that we deployed locally. We also implemented RAG querying by integrating LlamaIndex with both OpenAI and Ollama APIs. 

\review{In our experiments, we set temperature = 2.0 for \approach and temperature = 0.75 for the DeepSeek R1 and Llama3 families, because we obtained the best F1-score with these values, as shown in Figure \ref{fig:inverse-scaling}. 
We set top\_p = 1 as OpenAI suggests for GPT models,\footnote{\href{https://platform.openai.com/docs/api-reference/responses/create\#responses-create-temperature}{https://platform.openai.com/docs/api-reference/responses/create}} and top\_p = 0.9, top\_k = 40 for DeepSeek R1 and Llama3 families, as the Ollama documentation suggests.\footnote{\href{https://github.com/ollama/ollama/blob/main/docs/api.md}{https://github.com/ollama/docs/api.md}} We ran each experiment three times and report the average scores. Three repetitions are widely used in the software engineering literature as a pragmatic compromise between statistical reliability of the results and computational cost~\cite{xue2025, yan-etal-2023-codetransocean}.
}

The fine-tuning process of \approach took 684 seconds (11 minutes) with OpenAI fine-tuning API. 
The RAG indexing of the complete Java code base  took on average 0.24 seconds per 1,000 lines of code when experimented on our dataset. 
Both fine-tuning and RAG indexing are one time effort only, and both are reasonably efficient.

\subsubsection*{Metrics} 

We evaluate the classification in terms of precision, recall, and F1-score for each class of scenarios (\tested, \untested, and \errorProne). 
We  compute precision and recall with respect to the ground truth. 
We compute the score for a given class (for instance \tested), by considering the other two classes (for instance \untested and \errorProne) as negative samples. 
We compute the F1-score across all the three predicted classes, that is, the average of the three F1-scores associated with each class. All these metrics have been computed with the scikit-learn library.\footnote{\url{https://scikit-learn.org/}} We also record the inference time and consumed tokens of all examined LLMs.

\medskip

Table~\ref{tab:rq1} presents the results of our experiments that address \emph{RQ1} and \emph{RQ2}.  The Table reports precision (\emph{P}), recall (\emph{R}), and F1-score (\emph{F1}) for each class of scenarios, \tested (columns \Tested), \untested, (columns \Untested) and \errorProne (columns \ErrorProne). Columns \emph{Not-yet-tested} groups \untested, and \errorProne scenarios, and presents the average F1-score for the two classes that are relevant for testing (Column \emph{Not-yet-tested/Avg. F1}).  The last column (\emph{Total Avg. F1}) reports the average F1-score for the whole set of scenarios.

\subsection{RQ1: Impact of LLMs}
\label{sec:llm}

RQ1 investigates the impact of LLMs on \approach.  
The \emph{RQ1} rows (label \emph{RQ1} in the first column) report the results of experimenting with vanilla LLMs (label \emph{Vanilla} in the second column), few-shot learning (label \emph{Few-shot learning}) and RAG (label \emph{RAG}).  
The vanilla experiments classify the input scenarios by instantiating the LLM of \classifier with \review{GPT family (3.5Turbo, 4Turbo, 4o), DeepSeek R1 family (1.5B, 7B, 14B, 32B and 70B) and Llama3 family (1B, 3B, 8B, and 70B)}. The results are reported in the rows with the corresponding labels in the third column of the table. 
The experiments with \gptfturbo augmented with combinations of 3, 6 and 9 shots as the LLM of the \classifier are reported in the rows labeled \emph{Few-shot learning} in the second column.
The experiments of the LLM with RAG are reported in the \emph{RAG} rows of Table~\ref{tab:rq1}.  
The last row of the \emph{RQ1} section of the table (label \emph{Baseline} in the second column) reports the results we obtained with a random classifier (i.e., each type of scenarios has equal probability), which we implemented with \texttt{scikit-learn} library that we executed ten independent times on the same dataset.

The results indicate that the vanilla LLMs  do not significantly improve over the random classifier. 
\review{Only two LLMs (Llama3 1B and Deepseek R1 14B) perform slightly better than the random classifier for \tested scenarios.}
Some LLMs perform better than the random classifier  for the \emph{Not-yet-tested} scenarios, but the total average F1-score of the LLMs is consistently lower than the baseline \review{except for Deepseek R1 14B (+1\%)}.  Overall, \llamaeight and  \gpt perform equally well in identifying not-yet-tested scenarios.

The \emph{Few-shot learning} rows of the table report the results of augmenting \gptfturbo with few-shot learning, which does not improve the results, and we did not invest further effort on few-shot learning.

\begin{figure}[th]
	\centering
	\includegraphics[width=.85\linewidth]{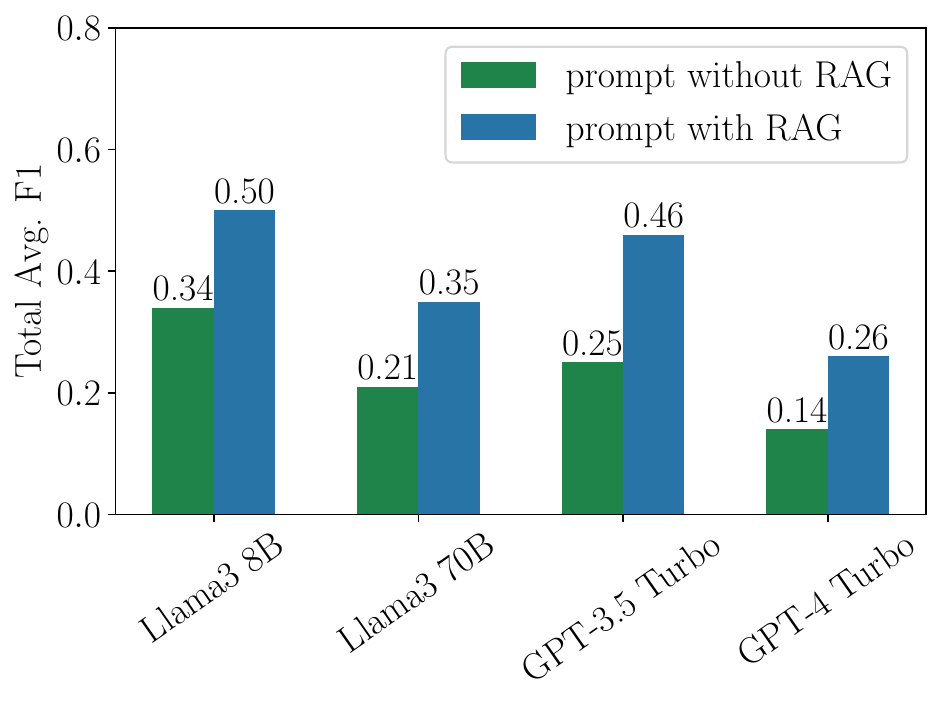}
	\caption{Total Avg. F1-score on large test suite for the LLMs}
	\label{fig:f1_lts}
\end{figure}

The \emph{RAG} rows of the table report the results of experimenting with RAG.  
The F1-scores indicate that RAG only marginally improves the results of the LLMs. RAG obtains the best results with \llamaeight with an F1-score that improves from 32\% to 39\%.  
We terminated the experiments with \gptfturbo, since the experiments with RAG are extremely expensive and we decided to stop experimenting, once we had evidence of the relatively small impact of RAG.   
The test suites of the scenarios in our dataset contain 68 test cases on average, and  
only 6.6\% scenarios of our dataset include test suites with more than 16K tokens, which largely exceed the context window of LLMs, and thus benefit from RAG. 
Therefore, the results with RAG may underestimate the contribution of RAG.
Figure~\ref{fig:f1_lts} compares the F1-scores of various LLMs prompted with and without RAG, computed only on scenarios with a large test suite that is larger than 16K tokens.  The diagrams in the figure show the relevance of RAG to scale up. The overall F1-scores gain at least a 12\% improvement with RAG.

The last two rows of the table (label \emph{Fine-tuning} in the second column) report the results of experimenting with \approach that fine tunes \gpt, one of the LLMs that performs the best according to our experimental  evaluation. 
The \approach fine-tuning of \gpt largely improves the total average F1-score, with an overall 80\% relative improvement, and significant gains for \tested (+63\%) and \errorProne (+25\%) scenarios. We argue that the relative drop of the F1-score for \untested scenarios is an acceptable consequence of the improved alignments of \approach for the scenarios of the other types. 
\approach largely improves over the random classifier as well, with a \review{66.6\%} relative improvement of the overall F1-score.

The results of \approach with and without RAG (rows \approachR and \approach, respectively) confirm the relatively small impact of RAG on the LLMs in this context.  \approachR works best for \emph{Not-yet-tested} scenarios (Avg. F1 53\%) and specifically for \untested scenarios (F1 56\%), while \approach works best for both \tested and \errorProne scenarios (F1 78\% and 53\%, respectively). 

The large improvement of \approach for \tested scenarios (78\% F1-score, more than double than both the random classifier and all LLMs) paired with the best F1-score for \errorProne scenarios indicate that \approach can indeed largely improve the test suite without a combinatorial explosion of the suite.     

\begin{figure}[t]
	\centering
	\begin{subfigure}[t]{\linewidth}
		\centering
	\includegraphics[width=.85\linewidth]{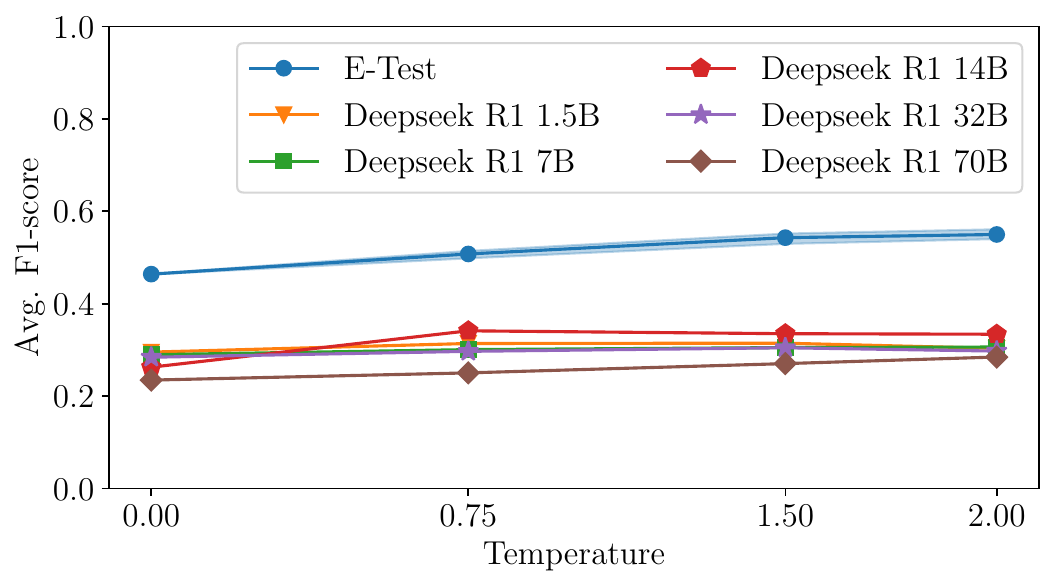}
	\caption{\review{Deepseek family}}
	\end{subfigure}
  \hfill
  \begin{subfigure}[t]{\linewidth}
	\centering
	\includegraphics[width=.85\linewidth]{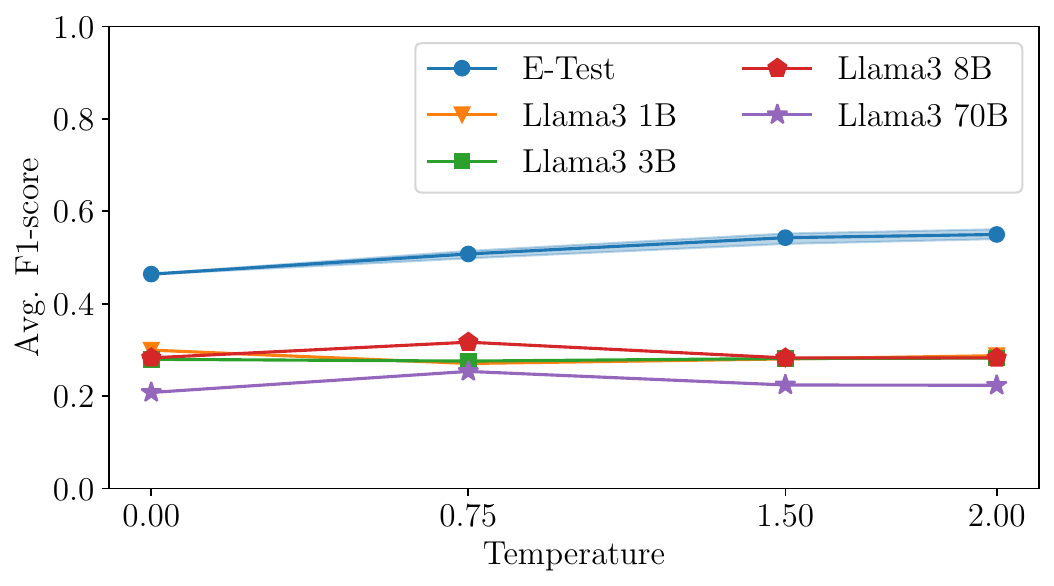}
	\caption{\review{Llama3 family}}
	\end{subfigure}
	\caption{\review{Inverse-scaling behaviour on \approach}}
	\label{fig:inverse-scaling}
\end{figure}

\review{We ran an inverse-scaling experiment on two families of models---Deepseek R1 (1.5B, 7B, 14B, 32B and 70B) and Llama3 (1B, 3B, 8B and 70B)---using different temperatures (0.0, 0.75, 1.50 and 2.00), with top\_p = 0.9 and top\_k = 40.
Figure \ref{fig:inverse-scaling} shows the average F1-score of each model with various temperatures.
\approach outperforms all LLM models with the best F1-score of \(0.55 \pm 0.01\) over 3 repetitions with temperature = 2. \approach also maintained an average F1-score of \(0.52 \pm 0.03\) across 4 various temperatures.
The slight variance of the average F1-score of \approach (the tiny shadow around the diagram in Figure~\ref{fig:inverse-scaling}) shows the stable performance of \approach.}

\review{The experiments indicate that \emph{smaller} LLMs outperform large models (70B) for all temperatures.
The Deepseek R1 14B model achieves the best F1-score of \(0.34 \pm 0.01\) with temperature = 0.75 higher than the best F1-score of \(0.28 \pm 0.05\) with temperature = 2 of the 70B model.
Similarly, the Llama3 8B model achieves the best F1-score of \(0.32 \pm 0.00\) with temperature = 0.75 higher than the best F1-score of \(0.25 \pm 0.00\) with temperature = 2 of the 70B model.
We argue that the results may depend on the frequencies of the types of test cases in the training set, the phenomenon that McKenzie et al. indicate as ``unwanted imitation''\cite{mckenzie2023inverse}: there may be a higher percentage of \textit{already-tested} than \textit{not-yet-tested} in large training sets, and this may impact the performance of the models.
}

\begin{summary}
	~\textbf{RQ1 Findings}: The LLM does have an impact on \approach. \approach (fine-tuned \gpt) outperforms the random classifier baseline, and performs best among the LLMs we evaluated for this task, with potential improvements with RAG for large test suites.
\end{summary}

\subsection{RQ2: Comparative Evaluation}
\label{sec:effectiveness}

RQ2 investigates the improvement of \approach over the state of the art. We compare \approach  against techniques for improving the quality of a test suite: \textit{FAST++} and \textit{field-ready testing}. 

\textit{FAST++} reduces the size of a regression test suite without executing the test cases: It converts the test cases of the suite to vectors with Term Frequency-Inverse Document Frequency (TF-IDF), clusters the vectors with K-means, and reduces the suites to the test cases in the $K$ centers of the clusters.
We classified all the scenarios $\langle In, M, T \rangle$ in our dataset, as \tested, \untested and \errorProne using the inverse order of the prioritization of \textit{FAST++}.  Row~\citet{Cruciani:ScalableApproachesForTestSuiteReduction:ICSE:2019} in block \emph{RQ2} of the table reports the results. 
\approach outperforms \textit{FAST++} for all indexes, but for the recall of \untested, where  \textit{FAST++} performs only slightly better (57\% over 54\%). The large gap of the Total Avg. F1 (50\% and 54\% for \approachR and \approach over 34\% for  \textit{FAST++}) confirms the improvement of \approach over \textit{FAST++}. 

The \textit{field-ready testing} approach of \citet{gazzola2022} generates test cases from scenarios that emerge in production. It selects the scenarios that deserve further testing with a grammar-based mechanism, and instantiates parametric test case templates for the selected scenarios. 
We experimented with the trigger that \cite{gazzola2022} report as best performing (1.4\%). We classified the scenarios that \textit{field-ready testing} selects from the Defects4J dataset  to generate test cases as \errorProne, and the scenarios that \textit{field-ready testing}  does not select as \tested.  
The \textit{field-ready testing} approach of Gazzola et al. does not allow distinguishing \untested from \errorProne scenarios.
\approach outperforms the \textit{field-ready testing} approach for all indexes, but for the recall of \untested, where \textit{field-ready testing} does not miss any \tested scenarios, due to the very low optimal trigger that \cite{gazzola2022} choses to minimize the probability of missing interesting test cases. The 94\% result of \approach is only slightly worse.  The large gap of the Total Avg. F1 (50\% and 54\% for \approachR and \approach over 30\% for  \textit{field-ready testing}) confirms the improvement of \approach over \textit{field-ready testing}.

\review{The F1-score of 0.55 that we obtain for a three-way classification problem is indeed good, since the main literature agrees on considering a F1-score above 0.5 sufficient for less demanding binary classification problems~\cite{buhl2023f1, hinojosa2024performance, opitz2019macro}.  
The comparison to the random baseline (F1-score 0.33), the best rule-based approach (F1-score 0.30), the best learning-based approach (F1-score 0.34), and the best LLM approach (F1-score 0.39) further confirms the quality of the 0.55 F1-score of \approach.}

\begin{summary}
	~\textbf{RQ2 Findings}: \approach performs significantly better than \textit{FAST++} (\review{+61.8\%} relative improvement) and \textit{field-ready testing} (\review{+83.3\%} relative improvement), which are the approaches functionally closest to \approach, in terms of total average F1-score.
\end{summary}

\subsection{RQ3: Impact of Queries}
\label{sec:queries}

\begin{figure}[t]
	\centering
	\includegraphics[width=.85\linewidth]{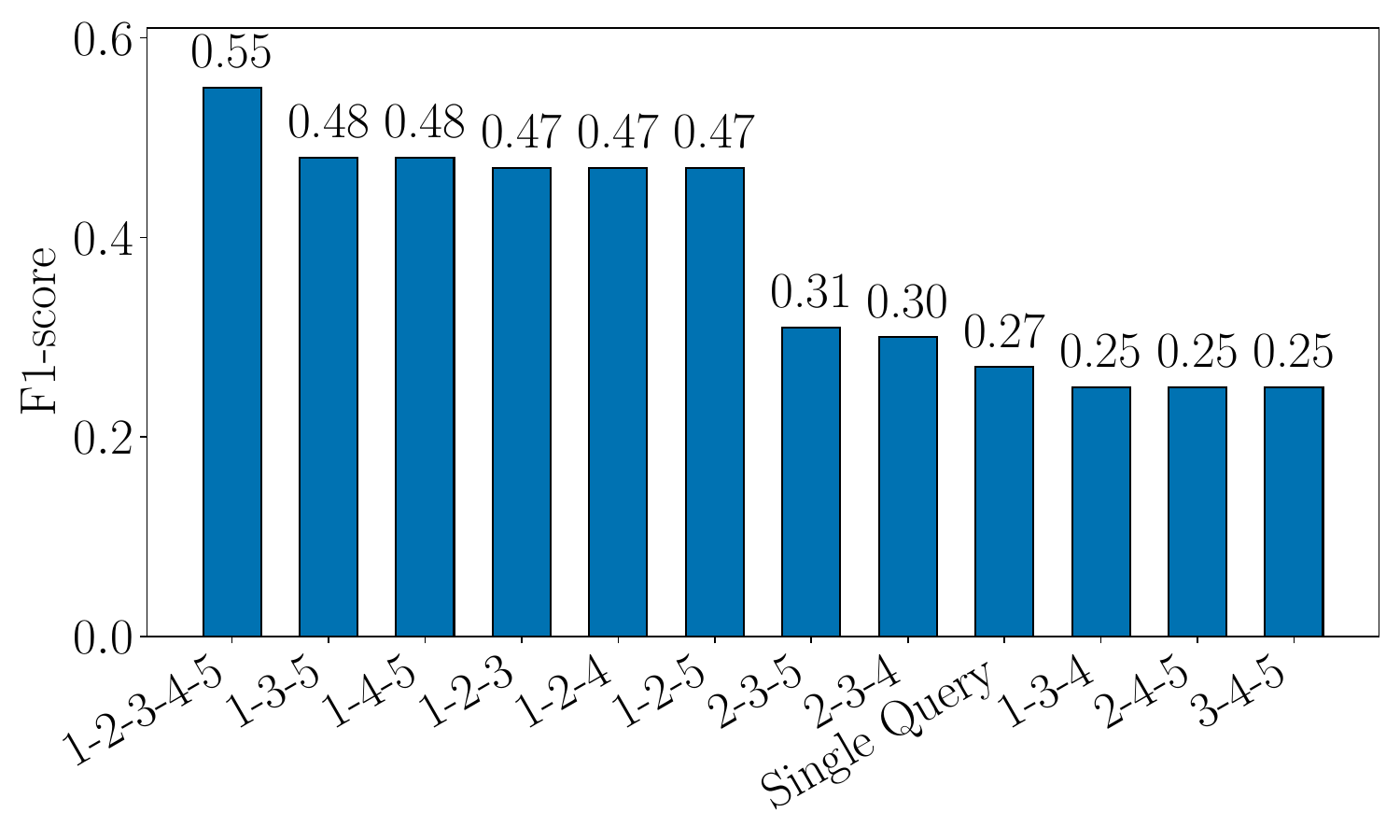}
	\caption{Impact of Query combinations on F1-score (1-2-3 means a combination of Q1, Q2 and Q3 in order)}
	\label{fig:query_combination}
\end{figure}

RQ3 investigates the impact of the combinations  of queries in the prompt of \approach. 
Figure~\ref{fig:query_combination} shows the F1-scores for all odd combinations of queries (the label of the x-axis), sorted by F1-score.  
The label \emph{1-2-3-4-5} indicates the five queries of \approach, the label \emph{Single~Query} indicates the single query \emph{“Given \mut, \mutscenario and \mutsuite, you should answer a number indicating its priority for testing. Answer 1 if \mutscenario is already-tested. Answer 2 if \mutscenario is need-test. Answer 3 if \mutscenario is error-prone. Do NOT explain your answer”}. 
We experimented with only combinations of three queries, since we need (i) at least two queries for a ternary classification, (ii) an odd number of queries to break ties, and (iii) a limited number of queries to keep the size of the prompt within a reasonable limit. 
We added the single query to study the need of multiple queries.   

The experiment confirms the better performance of five queries over subsets of three queries. It indicates a nonevent performance of different combinations of three queries, due to the different dependencies among queries. It indicates that five queries clearly outperform a single query, while three queries only do not always perform better than a single query.  

\begin{summary}
	~\textbf{RQ3 Findings}: The combinations of queries do significantly impact the classification, with the whole set of five queries performing significantly better than any subset of queries, while keeping the prompt within a reasonable size.
\end{summary}

\subsection{RQ4: Efficiency}

\begin{figure}[th]
	\centering
	\begin{subfigure}[t]{\linewidth}
		\centering
		\includegraphics[width=.85\linewidth]{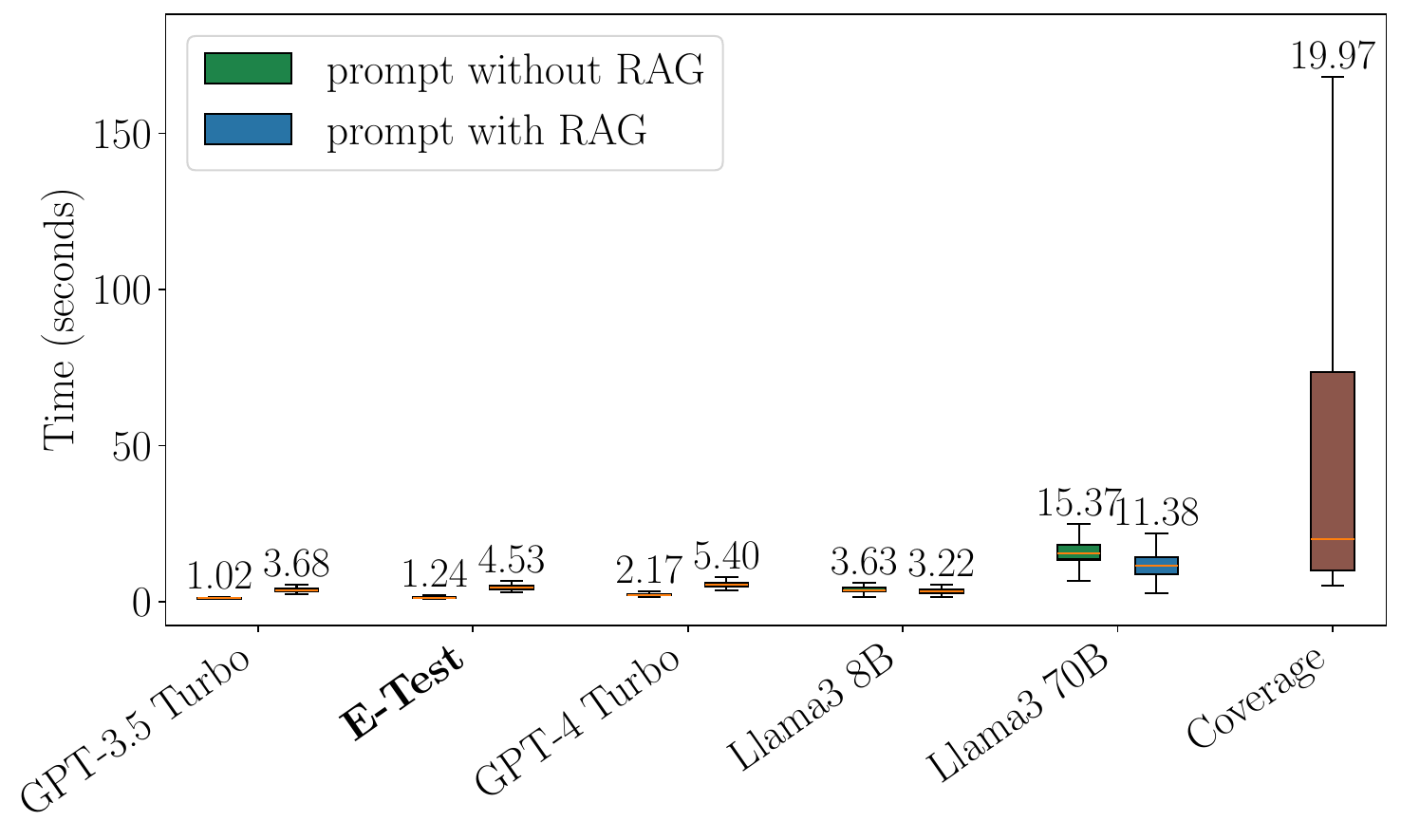}
		\caption{Query time of different models}
		\label{fig:time_boxplot}
	\end{subfigure}
	
	\begin{subfigure}[t]{\linewidth}
		\centering
		\includegraphics[width=.85\linewidth]{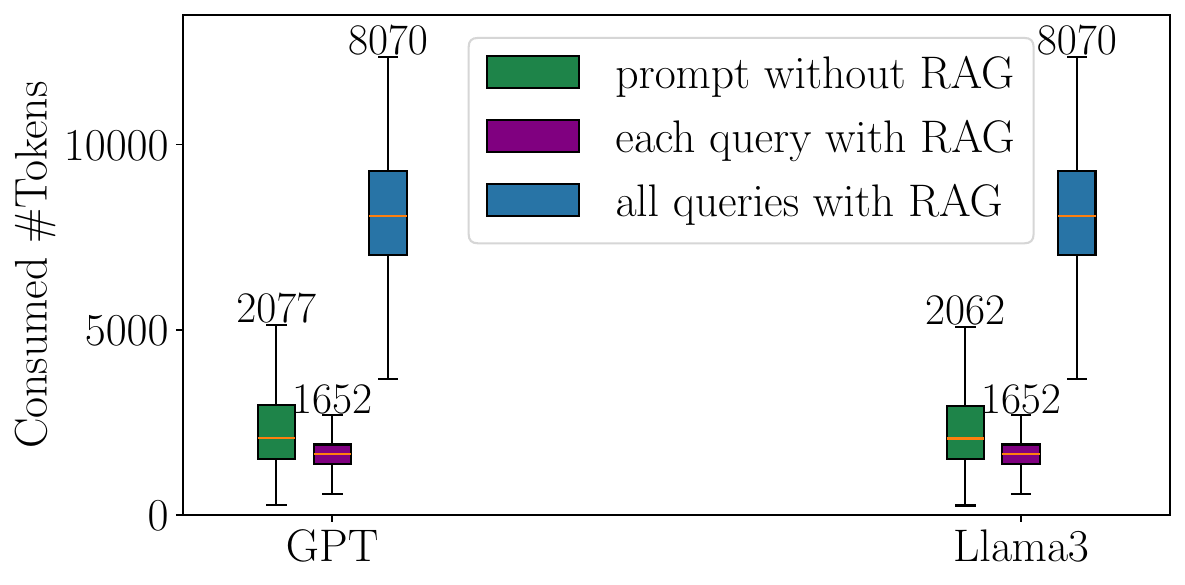}
		\caption{Token consumption for the queries}
		\label{fig:tokens_boxplot}
	\end{subfigure}
	\caption{Query time and token consumption for different configurations (median numbers are at the top of box plots) }
	\label{fig:scalability_figures}
\end{figure}

The execution time of  LLMs to classify the scenarios is the main factors that limits the practical applicability of \approach.  RQ4 investigates the efficiency of \approach, by measuring both the response time and the token consumption of the LLMs, the potential bottlenecks for the efficiency of \approach.  We measure the response time as the time that \approach takes to classify the scenarios with the different LLMs, and we compare the results with the time that \approach takes to select \emph{not-yet-tested} scenarios by simply executing the scenarios and determining the increment of branch coverage with respect to the test suite. In this way, we quantify the gain of LLMs with respect to the equivalence of the scenarios according to branch coverage. 
We collected the inference time for each prompt with the OpenAI APIs and Ollama APIs for the examined models.

The plots in Figure~\ref{fig:time_boxplot} clearly show that all LLMs (first 5 pairs of plots) perform dramatically better than the direct computation of branch coverage (the last plot in the figure).   The figure indicates similar results for all LLMs, with slightly better results of the GPT family with respect to the Llama family, possibly due to both the different training of the models that we used for the experiments and the runtime setup of the experiments (we ran Llama3 models on our powerful cluster and GPT with  OpenAPIs\footnote{The two environments offer the similar GPUs, according to the publicly available information from OpenAI~\cite{Brown:GPT:NIPS:2020}}).   
The data in the figure indicate a high variance of computing branch coverage, with a wide range of values, from $5.12$ to $168.14$ seconds.
On the other hand, the response time of all LLMs remains within small boundaries, within a range between $0.81$ and $24.97$ seconds, in the worst case.
The data with and without RAG in the figure indicate the low impact of RAG on the computation time.

Figure~\ref{fig:tokens_boxplot} compares the token consumption for prompting with RAG for all five queries as a whole (blue), with RAG separately for each query at once (purple) and without RAG (green).  
The plots indicate no significant differences between GPT and Llama tokenizers. They also show a large impact of RAG:  Prompting queries with RAG (blue and purple\footnote{The purple plots in the figure report the tokens consumed for each single query with RAG, which is limited by the context window of LLMs. The consumption for five queries is five times the consumption for a single query reported in the plots.}
 plots in the figure) consumes more tokens than prompting without RAG (green plots). 
The lower variance of prompting with RAG separately for each query at once (purple) than without RAG (green) suggests that  
the token consumption of suitably prompting with RAG grows less than prompting without RAG, and thus improves the efficiency of \approach.  

\review{Querying LLMs about test coverage is less precise than directly executing the test cases.  However, querying LLMs about the possibility that a test case increases the coverage of a large test suite is far less expensive than executing the test cases in the suite in all contexts. 
Our experiments indicate that querying LLMs is an excellent trade-off between the high and  impractical costs of executing very large test suites and the precision of the response.
}

\begin{summary}
	~\textbf{RQ4 Findings}: \approach classifies scenarios efficiently in terms of response time with respect to the direct computation of branch coverage. The limited variance of consumed tokens further confirms the efficiency of \approach.  
\end{summary}

\subsection{RQ5: Test Case Generation}
\label{sec:rq_test_case_generation}

RQ5 investigates the ability of \approach to generate useful test cases.  
We generated test cases for the 673 bugs of Defects4J with accessible test cases that trigger the failures (trigger tests) and that we use as ground truth: We fed \approach with the scenarios that correspond to the trigger tests, and the test suites of the methods without the trigger tests.
We measured the usefulness of the generated test cases as the percentage of generated test cases that are syntactically correct, compile, reveal failures, and obtain the same or higher branch coverage than the ground truth. 
We  performed syntax check, compilation check, and failure check, with up to five retries for fixing syntax errors, compilation errors, and assertion failures for each sequence, as shown in Algorithm~\ref{alg:utcg}.

\begin{algorithm}[t]
	\small
	\caption{Unit Test Case Generation}
	\label{alg:utcg}
	\begin{algorithmic}[1]
	\Require execution scenario $\langle$\mutscenario, \mut, \mutsuite$\rangle$
	\Ensure unit test case $UTC$ \textbf{or} None
\State Execute \approach scenario classification
\State $UTC \gets$ Continue prompting with "Use \mutscenario to generate only one test case with an assertion for \mut but do not catch exceptions."
\State $MAX\_RETRY \gets 5$
\State $retry_{syntax}$, $retry_{compile}$, $retry_{assert} \gets 0$
\While{$retry_{syntax}$, $retry_{compile}$, $retry_{assert} < MAX\_RETRY$}
\State Parse $UTC$ with Python library javalang
\If{Syntax error in $UTC$}
	\State $retry_{syntax} \gets retry_{syntax} + 1$
	\State $UTC \gets$ Continue prompting with JavaSyntaxError message
\Else
	\State Compile \mutsuite with $UTC$
	\If{Compilation error in $UTC$}
	\State $retry_{compile} \gets retry_{compile} + 1$
	\State $UTC \gets$ Continue prompting with compilation error
	\Else
	\State Execute $UTC$
	\If{$UTC$ does not trigger failures}
	\State $retry_{assert} \gets retry_{assert} + 1$
	\State $UTC \gets$ Continue prompting to correct assertions
	\Else
	\State \Return $UTC$
	\EndIf
	\EndIf
	
\EndIf
\EndWhile
\State \Return None

	\end{algorithmic}
	\end{algorithm}

\begin{figure}[t]
	\centering
	\includegraphics[width=.85\linewidth]{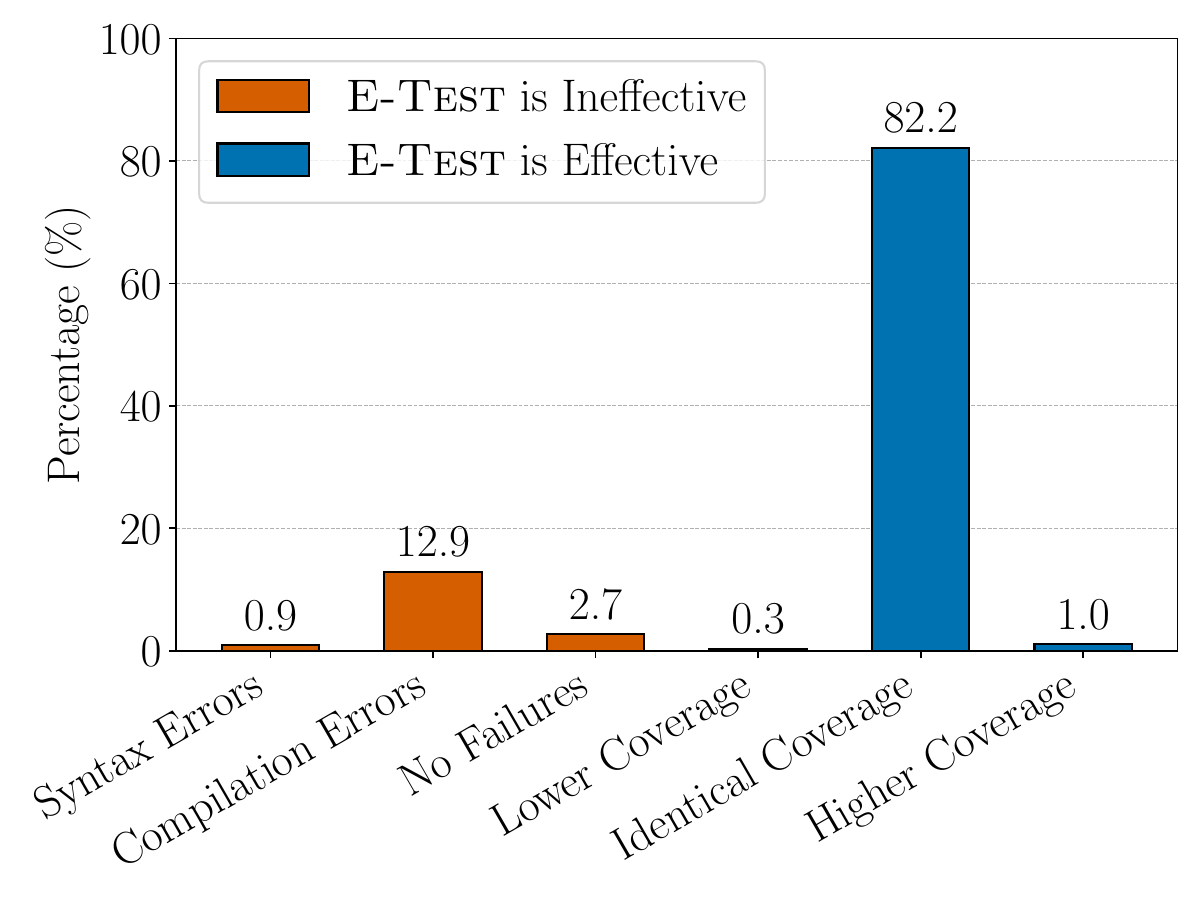}
	\caption{Test generation of \approach for Defects4J's 673 bugs}
	\label{fig:rq5-results}
\end{figure}

Figure~\ref{fig:rq5-results} summarizes the results of the experiment.
667 out of 673 generated tests (99.1\%) are syntactically correct (line 6 of Algorithm~\ref{alg:utcg}).
23 out of 667 tests required retries to fix some syntax errors, with a mean of 2 retries. 
580 out of 667 syntactically correct tests (87\% with respect to the syntactically correct tests and 86\% with respect to generated tests) successfully compile.
151 out of 580 tests required retries during compilation, with a mean of 1.8 retries. 
562 out of 580 compiled tests (97\% with respect to the compiled tests and 83.5\% with respect to generated tests) trigger the failure in the ground truth. 
85 tests out of 562 required retries to fix assertion issues, with a mean of 1 retry.
560 tests out of 562 tests  (99.5\% with respect to the failure-revealing tests and 83\% with respect to generated tests) reveal the failure with a branch coverage equal or even higher than the ground truth. 
Overall, \approach generates test cases that reveal the failures for 83.2\% failures in Defects4J.

\review{The results of our experiments indicate that \approach generates test cases that largely augment the coverage of the code and reveal hidden bugs.   The systematic usage of \approach on the scenarios observed in production can 
reduce both the gap between the code covered with the test suites and the code executed in production, and the number of bugs that escape the test and occur in production.}

\begin{summary}
	~\textbf{RQ5 Findings}: \approach generates a high number of useful test cases, that reveal failures by exercising buggy code.  
\end{summary}
\subsection{Threats to Validity}
\paragraph{\textit{Internal Validity}}

The main threat to the internal validity of our results comes from the heterogeneous hardware we use in the experiments (our cluster for Llama and the OpenAI API for GPT.)    
We mitigate the threat by configuring our cluster with the similar NVIDIA A100 GPUs that OpenAI offers, according to the available documentation. 
Discrepancies in the hardware configurations can change the results in terms of efficiency (RQ4) but do not impact the validity of \approach, as an approach that relies on an LLM core. 

\paragraph{\textit{External Validity}}

The main threat to the external validity of our results comes from the dataset. 
Defects4J is a well know dataset that is likely part of the general training set of LLMs.  
We mitigate the threat by augmenting Defects4J with scenarios we mine from recent Github projects that are less likely used for training LLMs, for their only recent availability on the Web.  
We also make the whole dataset we use in the experiments publicly available on a replication package to allow interested readers to replicate the experiments.

\section{Related Work}
\label{sec:related}

\approach improves the quality of test suites by sifting candidate test cases.  Here we briefly discuss test suite augmentation for regression testing and field testing approaches, the closest work that attempts to improve test suites with new test cases and mine new test cases from production, respectively, and we briefly overview the LLM-based approaches to generate test cases that are closest to \approach.

\paragraph{Test Suite Augmentation}

Test suite augmentation approaches generate test cases to reveal changed behaviors between different versions of the program~\cite{
    Xu:DirectedTestSuiteAugmentationTechniquesAndTradeoffs:FSE:2010,Zhang:JunitTestSuiteReduction:ISSRE:2011,
    Danglot:TestAmplification:JSS:2019, 
    Alshammari:FlakeFlagger:ICSE:2021,
    Liu:TestProgramMutation:ASE:2024}.
\citet{Santelices:TestSuiteAugmentation:ASE:2008}’s \textsc{MaTRIX} approach identifies inadequately-tested  behaviors with symbolic execution that analyzes the different chains and conditions up to a given distance around the changes, to contain the demands of resources of the technique. 
\citet{Bloem:AutomatingTestSuiteAugmentation:2014:SWQD}’s approach generates test cases that cover the branches related to functions that change across versions, by combining symbolic execution and model checking. 
\citet{Cruciani:ScalableApproachesForTestSuiteReduction:ICSE:2019}’s \textit{FAST++}~ statically selects representative regression tests from large-size test suites.  

\approach is complementary to test reduction approaches: Test suite reduction approaches optimize test suites, by filtering out redundant test cases,  \approach improves the quality of test suites, by adding test cases that improve the exploration of the runtime space. 

\paragraph{Field Testing}
Field testing approaches test software systems with scenarios that emerge in production~\cite{bertolino2021}. 
Most approaches focus on executing tests in production without disrupting the system~\cite{ceccato2020, moran2017,Porter:Skoll:TSE:2007}. \citet{murphy2009}’s in-vivo testing forks methods to execute them with runtime objects that it monitors from production to avoid interfering with the system behavior.
\citet{gazzola2022}’s field-ready testing  generates test cases from scenarios that emerge in production, by instantiating test templates.
\approach sifts useful test cases from scenarios with an LLM-core, and relies on field testing approaches to both mine scenarios from production and execute the test cases with the configurations from production. 

We discuss the experimental comparison of \approach with both Cruciani et al.’s \textit{FAST++} and Gazzola et al.’s \textit{field-ready testing}, the two closest approaches to \approach, in Section~\ref{sec:effectiveness}.

\paragraph{Large Language Models for Software Testing}
Large language models rely on attention mechanism of transformers to answer users' queries\cite{Yang:SparseCoder:ESE:2024, Chapman:LLMStaticAnalysis:SOAP:2024, Ouyang:ChatGPTCodeGeneration:TOSEM:2025}.
The strong capability of LLMs in understanding code-related tasks provides unique opportunities to analyze software without executing it~\cite{Pei:LLMProgramInvariants:ICML:2023, Nong:APPATCH:USENIX:2025}, especially for testing complex software systems~\cite{Fakhoury:LLMTestCodeGeneration:TSE:2024, wang2024, Fan:ProgramSelectionLLM:ISSTA:2024, Lu:DiaVio:ISSTA:2024, Wang:TESTEVAL:2025}.
\citet{Yang:LLMUnitTestGeneration:ASE:2024} conduct an empirical study about the capability of various open-source LLMs on generating unit test cases and highlight the importance of prompt design and fine-tuning for the task.
\citet{Kang:FewShotTesters:ICSE:2023}'s \textsc{Libro} generates bug-reproducing test cases via querying an LLM and augments the test suite with the top test case according to a heuristics-based ranking.
\citet{Su:ExploratoryTestingLLM:ICSE:2024}'s SysKG-UTF generates testing scenarios from bug reports via constructing knowledge graph with LLMs.

We use a fine-tuned LLM to classify execution scenarios using a well-structured prompt template and relevant context with RAG.
\section{Conclusion}
\label{sec:conclusion}

In this paper, we propose \emph{e’er-improving testing}, and present \approach. 
\emph{E’er-improving testing} is a new viewpoint on long lasting testing that upsets the perspective of testing: We can automatically  improve test suites by sifting massive amounts of scenarios that become available from many sources, notably the production environment and the automatic generation of large sets of test cases. 
 \approach fine tunes an LLM core and engineers prompts to gain the information relevant to separate \tested, \untested and \errorProne scenarios of Java methods. It uses the information to identify scenarios that can improve the test suite, namely  \untested and \errorProne scenarios, and generates test cases from them.
The results of the empirical evaluation show that \approach can effectively identify not-yet-tested scenarios and generate error-revealing test cases.  The comparison with the ground truth and the closest related approaches confirm the improvement of \approach over the state of the art.    
In a nutshell, our research contributes to advancing automated software testing methodologies and highlights the potential of LLMs in improving software reliability with an impact for both researchers and developers.

We are currently integrating our approach into system testing pipelines to enhance end-to-end testing efficiency and robustness~\cite{Brunetto:AutomaticTestCaseGeneration:JSS:2021}.
\review{We are studying both long context LLMs (context window $>$ 1M tokens) extended with advanced RAG techniques and extensions of the prompt template with queries that can examine scenarios beyond inputs, such as telemetry data (logs, metrics, and traces) to address the core issue of scaling \approach from unit to integration and system testing, namely the high-quality context.}

\begin{acks}
  This work is supported by the \grantsponsor{SNF}{Swiss SNF project A-Test Autonomic Software Testing}{} ({SNF 200021\_215487}).

\end{acks}

\balance
\bibliographystyle{ACM-Reference-Format}
\bibliography{references}


\begin{thebibliography}{69}


\ifx \showCODEN    \undefined \def \showCODEN     #1{\unskip}     \fi
\ifx \showDOI      \undefined \def \showDOI       #1{#1}\fi
\ifx \showISBNx    \undefined \def \showISBNx     #1{\unskip}     \fi
\ifx \showISBNxiii \undefined \def \showISBNxiii  #1{\unskip}     \fi
\ifx \showISSN     \undefined \def \showISSN      #1{\unskip}     \fi
\ifx \showLCCN     \undefined \def \showLCCN      #1{\unskip}     \fi
\ifx \shownote     \undefined \def \shownote      #1{#1}          \fi
\ifx \showarticletitle \undefined \def \showarticletitle #1{#1}   \fi
\ifx \showURL      \undefined \def \showURL       {\relax}        \fi
\providecommand\bibfield[2]{#2}
\providecommand\bibinfo[2]{#2}
\providecommand\natexlab[1]{#1}
\providecommand\showeprint[2][]{arXiv:#2}

\bibitem[Alshahwan et~al\mbox{.}(2024)]%
        {Alshahwan:AutomatedUnitTestImprovement:FSE:2024}
\bibfield{author}{\bibinfo{person}{Nadia Alshahwan}, \bibinfo{person}{Jubin Chheda}, \bibinfo{person}{Anastasia Finogenova}, \bibinfo{person}{Beliz Gokkaya}, \bibinfo{person}{Mark Harman}, \bibinfo{person}{Inna Harper}, \bibinfo{person}{Alexandru Marginean}, \bibinfo{person}{Shubho Sengupta}, {and} \bibinfo{person}{Eddy Wang}.} \bibinfo{year}{2024}\natexlab{}.
\newblock \showarticletitle{Automated Unit Test Improvement using Large Language Models at Meta}. In \bibinfo{booktitle}{\emph{Companion Proceedings of the 32nd ACM International Conference on the Foundations of Software Engineering}} (Porto de Galinhas, Brazil) \emph{(\bibinfo{series}{FSE 2024})}. \bibinfo{publisher}{Association for Computing Machinery}, \bibinfo{address}{New York, NY, USA}, \bibinfo{pages}{185–196}.
\newblock
\showISBNx{9798400706585}
\urldef\tempurl%
\url{https://doi.org/10.1145/3663529.3663839}
\showDOI{\tempurl}


\bibitem[Alshammari et~al\mbox{.}(2021)]%
        {Alshammari:FlakeFlagger:ICSE:2021}
\bibfield{author}{\bibinfo{person}{Abdulrahman Alshammari}, \bibinfo{person}{Christopher Morris}, \bibinfo{person}{Michael Hilton}, {and} \bibinfo{person}{Jonathan Bell}.} \bibinfo{year}{2021}\natexlab{}.
\newblock \showarticletitle{FlakeFlagger: Predicting Flakiness Without Rerunning Tests}. In \bibinfo{booktitle}{\emph{Proceedings of the 43rd International Conference on Software Engineering}} (Madrid, Spain) \emph{(\bibinfo{series}{ICSE '21})}. \bibinfo{publisher}{IEEE Press}, \bibinfo{pages}{1572–1584}.
\newblock
\showISBNx{9781450390859}
\urldef\tempurl%
\url{https://doi.org/10.1109/ICSE43902.2021.00140}
\showDOI{\tempurl}


\bibitem[Aniche et~al\mbox{.}(2022)]%
        {Zaidman:EngineerTestCases:TSE:2022}
\bibfield{author}{\bibinfo{person}{Maurício Aniche}, \bibinfo{person}{Christoph Treude}, {and} \bibinfo{person}{Andy Zaidman}.} \bibinfo{year}{2022}\natexlab{}.
\newblock \showarticletitle{How Developers Engineer Test Cases: An Observational Study}.
\newblock \bibinfo{journal}{\emph{IEEE Transactions on Software Engineering}} \bibinfo{volume}{48}, \bibinfo{number}{12} (\bibinfo{year}{2022}), \bibinfo{pages}{4925--4946}.
\newblock
\urldef\tempurl%
\url{https://doi.org/10.1109/TSE.2021.3129889}
\showDOI{\tempurl}


\bibitem[Bertolino et~al\mbox{.}(2021)]%
        {bertolino2021}
\bibfield{author}{\bibinfo{person}{Antonia Bertolino}, \bibinfo{person}{Pietro Braione}, \bibinfo{person}{Guglielmo~De Angelis}, \bibinfo{person}{Luca Gazzola}, \bibinfo{person}{Fitsum Kifetew}, \bibinfo{person}{Leonardo Mariani}, \bibinfo{person}{Matteo Orr\`{u}}, \bibinfo{person}{Mauro Pezz\`{e}}, \bibinfo{person}{Roberto Pietrantuono}, \bibinfo{person}{Stefano Russo}, {and} \bibinfo{person}{Paolo Tonella}.} \bibinfo{year}{2021}\natexlab{}.
\newblock \showarticletitle{A Survey of Field-based Testing Techniques}.
\newblock \bibinfo{journal}{\emph{ACM Comput. Surv.}} \bibinfo{volume}{54}, \bibinfo{number}{5}, Article \bibinfo{articleno}{92} (\bibinfo{date}{May} \bibinfo{year}{2021}), \bibinfo{numpages}{39}~pages.
\newblock
\showISSN{0360-0300}
\urldef\tempurl%
\url{https://doi.org/10.1145/3447240}
\showDOI{\tempurl}


\bibitem[Bloem et~al\mbox{.}(2014)]%
        {Bloem:AutomatingTestSuiteAugmentation:2014:SWQD}
\bibfield{author}{\bibinfo{person}{Roderick Bloem}, \bibinfo{person}{Robert Koenighofer}, \bibinfo{person}{Franz Röck}, {and} \bibinfo{person}{Michael Tautschnig}.} \bibinfo{year}{2014}\natexlab{}.
\newblock \showarticletitle{Automating Test-Suite Augmentation}. In \bibinfo{booktitle}{\emph{2014 14th International Conference on Quality Software}}. \bibinfo{pages}{67--72}.
\newblock
\urldef\tempurl%
\url{https://doi.org/10.1109/QSIC.2014.40}
\showDOI{\tempurl}


\bibitem[Brown et~al\mbox{.}(2020)]%
        {Brown:GPT:NIPS:2020}
\bibfield{author}{\bibinfo{person}{Tom~B. Brown}, \bibinfo{person}{Benjamin Mann}, \bibinfo{person}{Nick Ryder}, \bibinfo{person}{Melanie Subbiah}, \bibinfo{person}{Jared Kaplan}, \bibinfo{person}{Prafulla Dhariwal}, \bibinfo{person}{Arvind Neelakantan}, \bibinfo{person}{Pranav Shyam}, \bibinfo{person}{Girish Sastry}, \bibinfo{person}{Amanda Askell}, \bibinfo{person}{Sandhini Agarwal}, \bibinfo{person}{Ariel Herbert-Voss}, \bibinfo{person}{Gretchen Krueger}, \bibinfo{person}{Tom Henighan}, \bibinfo{person}{Rewon Child}, \bibinfo{person}{Aditya Ramesh}, \bibinfo{person}{Daniel~M. Ziegler}, \bibinfo{person}{Jeffrey Wu}, \bibinfo{person}{Clemens Winter}, \bibinfo{person}{Christopher Hesse}, \bibinfo{person}{Mark Chen}, \bibinfo{person}{Eric Sigler}, \bibinfo{person}{Mateusz Litwin}, \bibinfo{person}{Scott Gray}, \bibinfo{person}{Benjamin Chess}, \bibinfo{person}{Jack Clark}, \bibinfo{person}{Christopher Berner}, \bibinfo{person}{Sam McCandlish}, \bibinfo{person}{Alec Radford}, \bibinfo{person}{Ilya Sutskever}, {and} \bibinfo{person}{Dario Amodei}.} \bibinfo{year}{2020}\natexlab{}.
\newblock \showarticletitle{Language models are few-shot learners}. In \bibinfo{booktitle}{\emph{Proceedings of the 34th International Conference on Neural Information Processing Systems}} (Vancouver, BC, Canada) \emph{(\bibinfo{series}{NIPS '20})}. \bibinfo{publisher}{Curran Associates Inc.}, \bibinfo{address}{Red Hook, NY, USA}, Article \bibinfo{articleno}{159}, \bibinfo{numpages}{25}~pages.
\newblock
\showISBNx{9781713829546}


\bibitem[Brunetto et~al\mbox{.}(2021)]%
        {Brunetto:AutomaticTestCaseGeneration:JSS:2021}
\bibfield{author}{\bibinfo{person}{Matteo Brunetto}, \bibinfo{person}{Giovanni Denaro}, \bibinfo{person}{Leonardo Mariani}, {and} \bibinfo{person}{Mauro Pezzè}.} \bibinfo{year}{2021}\natexlab{}.
\newblock \showarticletitle{On introducing automatic test case generation in practice: A success story and lessons learned}.
\newblock \bibinfo{journal}{\emph{Journal of Systems and Software}}  \bibinfo{volume}{176} (\bibinfo{year}{2021}), \bibinfo{pages}{110933}.
\newblock
\showISSN{0164-1212}
\urldef\tempurl%
\url{https://doi.org/10.1016/j.jss.2021.110933}
\showDOI{\tempurl}


\bibitem[Buhl(2023)]%
        {buhl2023f1}
\bibfield{author}{\bibinfo{person}{Nikolaj Buhl}.} \bibinfo{year}{2023}\natexlab{}.
\newblock \showarticletitle{F1 Score in Machine Learning}.
\newblock \bibinfo{journal}{\emph{Eri{\c{s}}im adresi: https://encord. com/blog/f1-score-in-machine-learning}} (\bibinfo{year}{2023}).
\newblock


\bibitem[Caldas et~al\mbox{.}(2024)]%
        {Caldas:FieldBasedTesting:TSE:2024}
\bibfield{author}{\bibinfo{person}{Ricardo Caldas}, \bibinfo{person}{Juan Antonio Pi\~{n}era Garc\'{\i}a}, \bibinfo{person}{Matei Schiopu}, \bibinfo{person}{Patrizio Pelliccione}, \bibinfo{person}{Gena\'{\i}na Rodrigues}, {and} \bibinfo{person}{Thorsten Berger}.} \bibinfo{year}{2024}\natexlab{}.
\newblock \showarticletitle{Runtime Verification and Field-Based Testing for ROS-Based Robotic Systems}.
\newblock \bibinfo{journal}{\emph{IEEE Trans. Softw. Eng.}} \bibinfo{volume}{50}, \bibinfo{number}{10} (\bibinfo{date}{Oct.} \bibinfo{year}{2024}), \bibinfo{pages}{2544–2567}.
\newblock
\showISSN{0098-5589}
\urldef\tempurl%
\url{https://doi.org/10.1109/TSE.2024.3444697}
\showDOI{\tempurl}


\bibitem[Ceccato et~al\mbox{.}(2020)]%
        {ceccato2020}
\bibfield{author}{\bibinfo{person}{Mariano Ceccato}, \bibinfo{person}{Davide Corradini}, \bibinfo{person}{Luca Gazzola}, \bibinfo{person}{Fitsum~Meshesha Kifetew}, \bibinfo{person}{Leonardo Mariani}, \bibinfo{person}{Matteo Orrù}, {and} \bibinfo{person}{Paolo Tonella}.} \bibinfo{year}{2020}\natexlab{}.
\newblock \showarticletitle{A Framework for In-Vivo Testing of Mobile Applications}. In \bibinfo{booktitle}{\emph{2020 IEEE 13th International Conference on Software Testing, Validation and Verification (ICST)}}. \bibinfo{pages}{286--296}.
\newblock
\urldef\tempurl%
\url{https://doi.org/10.1109/ICST46399.2020.00037}
\showDOI{\tempurl}


\bibitem[Chapman et~al\mbox{.}(2024)]%
        {Chapman:LLMStaticAnalysis:SOAP:2024}
\bibfield{author}{\bibinfo{person}{Patrick~J. Chapman}, \bibinfo{person}{Cindy Rubio-Gonz\'{a}lez}, {and} \bibinfo{person}{Aditya~V. Thakur}.} \bibinfo{year}{2024}\natexlab{}.
\newblock \showarticletitle{Interleaving Static Analysis and LLM Prompting}. In \bibinfo{booktitle}{\emph{Proceedings of the 13th ACM SIGPLAN International Workshop on the State Of the Art in Program Analysis}} (Copenhagen, Denmark) \emph{(\bibinfo{series}{SOAP 2024})}. \bibinfo{publisher}{Association for Computing Machinery}, \bibinfo{address}{New York, NY, USA}, \bibinfo{pages}{9–17}.
\newblock
\showISBNx{9798400706219}
\urldef\tempurl%
\url{https://doi.org/10.1145/3652588.3663317}
\showDOI{\tempurl}


\bibitem[Cheng et~al\mbox{.}(2024)]%
        {Cheng:TestCasePrioritization:ISSTA:2024}
\bibfield{author}{\bibinfo{person}{Runxiang Cheng}, \bibinfo{person}{Shuai Wang}, \bibinfo{person}{Reyhaneh Jabbarvand}, {and} \bibinfo{person}{Darko Marinov}.} \bibinfo{year}{2024}\natexlab{}.
\newblock \showarticletitle{Revisiting Test-Case Prioritization on Long-Running Test Suites}. In \bibinfo{booktitle}{\emph{Proceedings of the 33rd ACM SIGSOFT International Symposium on Software Testing and Analysis}} (Vienna, Austria) \emph{(\bibinfo{series}{ISSTA 2024})}. \bibinfo{publisher}{Association for Computing Machinery}, \bibinfo{address}{New York, NY, USA}, \bibinfo{pages}{615–627}.
\newblock
\showISBNx{9798400706127}
\urldef\tempurl%
\url{https://doi.org/10.1145/3650212.3680307}
\showDOI{\tempurl}


\bibitem[Chow et~al\mbox{.}(2024)]%
        {Chow2024}
\bibfield{author}{\bibinfo{person}{Yiu~Wai Chow}, \bibinfo{person}{Luca Di~Grazia}, {and} \bibinfo{person}{Michael Pradel}.} \bibinfo{year}{2024}\natexlab{}.
\newblock \showarticletitle{PyTy: Repairing Static Type Errors in Python}. In \bibinfo{booktitle}{\emph{Proceedings of the IEEE/ACM 46th International Conference on Software Engineering}} (Lisbon, Portugal) \emph{(\bibinfo{series}{ICSE '24})}. \bibinfo{publisher}{Association for Computing Machinery}, \bibinfo{address}{New York, NY, USA}, Article \bibinfo{articleno}{87}, \bibinfo{numpages}{13}~pages.
\newblock
\showISBNx{9798400702174}
\urldef\tempurl%
\url{https://doi.org/10.1145/3597503.3639184}
\showDOI{\tempurl}


\bibitem[Cruciani et~al\mbox{.}(2019)]%
        {Cruciani:ScalableApproachesForTestSuiteReduction:ICSE:2019}
\bibfield{author}{\bibinfo{person}{Emilio Cruciani}, \bibinfo{person}{Breno Miranda}, \bibinfo{person}{Roberto Verdecchia}, {and} \bibinfo{person}{Antonia Bertolino}.} \bibinfo{year}{2019}\natexlab{}.
\newblock \showarticletitle{Scalable Approaches for Test Suite Reduction}. In \bibinfo{booktitle}{\emph{2019 IEEE/ACM 41st International Conference on Software Engineering (ICSE)}}. \bibinfo{pages}{419--429}.
\newblock
\urldef\tempurl%
\url{https://doi.org/10.1109/ICSE.2019.00055}
\showDOI{\tempurl}


\bibitem[Danglot et~al\mbox{.}(2019)]%
        {Danglot:TestAmplification:JSS:2019}
\bibfield{author}{\bibinfo{person}{Benjamin Danglot}, \bibinfo{person}{Oscar Vera-Perez}, \bibinfo{person}{Zhongxing Yu}, \bibinfo{person}{Andy Zaidman}, \bibinfo{person}{Martin Monperrus}, {and} \bibinfo{person}{Benoit Baudry}.} \bibinfo{year}{2019}\natexlab{}.
\newblock \showarticletitle{A snowballing literature study on test amplification}.
\newblock \bibinfo{journal}{\emph{Journal of Systems and Software}}  \bibinfo{volume}{157} (\bibinfo{year}{2019}), \bibinfo{pages}{110398}.
\newblock
\showISSN{0164-1212}
\urldef\tempurl%
\url{https://doi.org/10.1016/j.jss.2019.110398}
\showDOI{\tempurl}


\bibitem[Elsner et~al\mbox{.}(2021)]%
        {Elsner2021}
\bibfield{author}{\bibinfo{person}{Daniel Elsner}, \bibinfo{person}{Florian Hauer}, \bibinfo{person}{Alexander Pretschner}, {and} \bibinfo{person}{Silke Reimer}.} \bibinfo{year}{2021}\natexlab{}.
\newblock \showarticletitle{Empirically evaluating readily available information for regression test optimization in continuous integration}. In \bibinfo{booktitle}{\emph{Proceedings of the 30th ACM SIGSOFT International Symposium on Software Testing and Analysis}} (Virtual, Denmark) \emph{(\bibinfo{series}{ISSTA 2021})}. \bibinfo{publisher}{Association for Computing Machinery}, \bibinfo{address}{New York, NY, USA}, \bibinfo{pages}{491–504}.
\newblock
\showISBNx{9781450384599}
\urldef\tempurl%
\url{https://doi.org/10.1145/3460319.3464834}
\showDOI{\tempurl}


\bibitem[Fakhoury et~al\mbox{.}(2024)]%
        {Fakhoury:LLMTestCodeGeneration:TSE:2024}
\bibfield{author}{\bibinfo{person}{Sarah Fakhoury}, \bibinfo{person}{Aaditya Naik}, \bibinfo{person}{Georgios Sakkas}, \bibinfo{person}{Saikat Chakraborty}, {and} \bibinfo{person}{Shuvendu~K. Lahiri}.} \bibinfo{year}{2024}\natexlab{}.
\newblock \showarticletitle{LLM-Based Test-Driven Interactive Code Generation: User Study and Empirical Evaluation}.
\newblock \bibinfo{journal}{\emph{IEEE Trans. Softw. Eng.}} \bibinfo{volume}{50}, \bibinfo{number}{9} (\bibinfo{date}{Sept.} \bibinfo{year}{2024}), \bibinfo{pages}{2254–2268}.
\newblock
\showISSN{0098-5589}
\urldef\tempurl%
\url{https://doi.org/10.1109/TSE.2024.3428972}
\showDOI{\tempurl}


\bibitem[Fan et~al\mbox{.}(2024)]%
        {Fan:ProgramSelectionLLM:ISSTA:2024}
\bibfield{author}{\bibinfo{person}{Zhiyu Fan}, \bibinfo{person}{Haifeng Ruan}, \bibinfo{person}{Sergey Mechtaev}, {and} \bibinfo{person}{Abhik Roychoudhury}.} \bibinfo{year}{2024}\natexlab{}.
\newblock \showarticletitle{Oracle-Guided Program Selection from Large Language Models}. In \bibinfo{booktitle}{\emph{Proceedings of the 33rd ACM SIGSOFT International Symposium on Software Testing and Analysis}} (Vienna, Austria) \emph{(\bibinfo{series}{ISSTA 2024})}. \bibinfo{publisher}{Association for Computing Machinery}, \bibinfo{address}{New York, NY, USA}, \bibinfo{pages}{628–640}.
\newblock
\showISBNx{9798400706127}
\urldef\tempurl%
\url{https://doi.org/10.1145/3650212.3680308}
\showDOI{\tempurl}


\bibitem[Fraser and Arcuri(2011)]%
        {fraser2011}
\bibfield{author}{\bibinfo{person}{Gordon Fraser} {and} \bibinfo{person}{Andrea Arcuri}.} \bibinfo{year}{2011}\natexlab{}.
\newblock \showarticletitle{EvoSuite: automatic test suite generation for object-oriented software}. In \bibinfo{booktitle}{\emph{Proceedings of the 19th ACM SIGSOFT Symposium and the 13th European Conference on Foundations of Software Engineering}} (Szeged, Hungary) \emph{(\bibinfo{series}{ESEC/FSE '11})}. \bibinfo{publisher}{Association for Computing Machinery}, \bibinfo{address}{New York, NY, USA}, \bibinfo{pages}{416–419}.
\newblock
\showISBNx{9781450304436}
\urldef\tempurl%
\url{https://doi.org/10.1145/2025113.2025179}
\showDOI{\tempurl}


\bibitem[Fraser and Arcuri(2012)]%
        {fraser2012whole}
\bibfield{author}{\bibinfo{person}{Gordon Fraser} {and} \bibinfo{person}{Andrea Arcuri}.} \bibinfo{year}{2012}\natexlab{}.
\newblock \showarticletitle{Whole test suite generation}.
\newblock \bibinfo{journal}{\emph{IEEE Transactions on Software Engineering}} \bibinfo{volume}{39}, \bibinfo{number}{2} (\bibinfo{year}{2012}), \bibinfo{pages}{276--291}.
\newblock


\bibitem[Gambi et~al\mbox{.}(2023)]%
        {Gambi2023}
\bibfield{author}{\bibinfo{person}{Alessio Gambi}, \bibinfo{person}{Hemant Gouni}, \bibinfo{person}{Daniel Berreiter}, \bibinfo{person}{Vsevolod Tymofyeyev}, {and} \bibinfo{person}{Mattia Fazzini}.} \bibinfo{year}{2023}\natexlab{}.
\newblock \showarticletitle{Action-Based Test Carving for Android Apps}. In \bibinfo{booktitle}{\emph{2023 IEEE International Conference on Software Testing, Verification and Validation Workshops (ICSTW)}}. \bibinfo{pages}{107--116}.
\newblock
\urldef\tempurl%
\url{https://doi.org/10.1109/ICSTW58534.2023.00032}
\showDOI{\tempurl}


\bibitem[Gazzola et~al\mbox{.}(2022)]%
        {gazzola2022}
\bibfield{author}{\bibinfo{person}{Luca Gazzola}, \bibinfo{person}{Leonardo Mariani}, \bibinfo{person}{Matteo Orrú}, \bibinfo{person}{Mauro Pezzè}, {and} \bibinfo{person}{Martin Tappler}.} \bibinfo{year}{2022}\natexlab{}.
\newblock \showarticletitle{Testing Software in Production Environments with Data from the Field}. In \bibinfo{booktitle}{\emph{2022 IEEE Conference on Software Testing, Verification and Validation (ICST)}}. \bibinfo{publisher}{IEEE Computer Society}, \bibinfo{pages}{58--69}.
\newblock
\urldef\tempurl%
\url{https://doi.org/10.1109/ICST53961.2022.00017}
\showDOI{\tempurl}


\bibitem[Gazzola et~al\mbox{.}(2017)]%
        {gazzola2017exploratory}
\bibfield{author}{\bibinfo{person}{Luca Gazzola}, \bibinfo{person}{Leonardo Mariani}, \bibinfo{person}{Fabrizio Pastore}, {and} \bibinfo{person}{Mauro Pezze}.} \bibinfo{year}{2017}\natexlab{}.
\newblock \showarticletitle{An exploratory study of field failures}. In \bibinfo{booktitle}{\emph{2017 IEEE 28th International Symposium on Software Reliability Engineering (ISSRE)}}. IEEE, \bibinfo{pages}{67--77}.
\newblock


\bibitem[Greca et~al\mbox{.}(2023)]%
        {greca2023}
\bibfield{author}{\bibinfo{person}{Renan Greca}, \bibinfo{person}{Breno Miranda}, {and} \bibinfo{person}{Antonia Bertolino}.} \bibinfo{year}{2023}\natexlab{}.
\newblock \showarticletitle{State of practical applicability of regression testing research: A live systematic literature review}.
\newblock \bibinfo{journal}{\emph{Comput. Surveys}} \bibinfo{volume}{55}, \bibinfo{number}{13s} (\bibinfo{year}{2023}), \bibinfo{pages}{1--36}.
\newblock


\bibitem[Harrold et~al\mbox{.}(1993)]%
        {Harrold:SizeOfTestSuite:TOSEM:1993}
\bibfield{author}{\bibinfo{person}{M.~Jean Harrold}, \bibinfo{person}{Rajiv Gupta}, {and} \bibinfo{person}{Mary~Lou Soffa}.} \bibinfo{year}{1993}\natexlab{}.
\newblock \showarticletitle{A methodology for controlling the size of a test suite}.
\newblock \bibinfo{journal}{\emph{ACM Trans. Softw. Eng. Methodol.}} \bibinfo{volume}{2}, \bibinfo{number}{3} (\bibinfo{date}{jul} \bibinfo{year}{1993}), \bibinfo{pages}{270–285}.
\newblock
\showISSN{1049-331X}
\urldef\tempurl%
\url{https://doi.org/10.1145/152388.152391}
\showDOI{\tempurl}


\bibitem[Hinojosa~Lee et~al\mbox{.}(2024)]%
        {hinojosa2024performance}
\bibfield{author}{\bibinfo{person}{Maria~Cristina Hinojosa~Lee}, \bibinfo{person}{Johan Braet}, {and} \bibinfo{person}{Johan Springael}.} \bibinfo{year}{2024}\natexlab{}.
\newblock \showarticletitle{Performance metrics for multilabel emotion classification: comparing micro, macro, and weighted f1-scores}.
\newblock \bibinfo{journal}{\emph{Applied Sciences}} \bibinfo{volume}{14}, \bibinfo{number}{21} (\bibinfo{year}{2024}), \bibinfo{pages}{9863}.
\newblock


\bibitem[Horgan et~al\mbox{.}(1994)]%
        {Horgan:Quality:Computer:1994}
\bibfield{author}{\bibinfo{person}{Joseph~R. Horgan}, \bibinfo{person}{Saul London}, {and} \bibinfo{person}{Michael~R. Lyu}.} \bibinfo{year}{1994}\natexlab{}.
\newblock \showarticletitle{Achieving software quality with testing coverage measures}.
\newblock \bibinfo{journal}{\emph{Computer}} \bibinfo{volume}{27}, \bibinfo{number}{9} (\bibinfo{date}{Sept.} \bibinfo{year}{1994}), \bibinfo{pages}{60–69}.
\newblock
\showISSN{0018-9162}
\urldef\tempurl%
\url{https://doi.org/10.1109/2.312032}
\showDOI{\tempurl}


\bibitem[Huang et~al\mbox{.}(2024)]%
        {Huang:TestCaseSelection:TOSEM:2024}
\bibfield{author}{\bibinfo{person}{Dong Huang}, \bibinfo{person}{Qingwen Bu}, \bibinfo{person}{Yichao Fu}, \bibinfo{person}{Yuhao Qing}, \bibinfo{person}{Xiaofei Xie}, \bibinfo{person}{Junjie Chen}, {and} \bibinfo{person}{Heming Cui}.} \bibinfo{year}{2024}\natexlab{}.
\newblock \showarticletitle{Neuron Sensitivity-Guided Test Case Selection}.
\newblock \bibinfo{journal}{\emph{ACM Trans. Softw. Eng. Methodol.}} \bibinfo{volume}{33}, \bibinfo{number}{7}, Article \bibinfo{articleno}{188} (\bibinfo{date}{Sept.} \bibinfo{year}{2024}), \bibinfo{numpages}{32}~pages.
\newblock
\showISSN{1049-331X}
\urldef\tempurl%
\url{https://doi.org/10.1145/3672454}
\showDOI{\tempurl}


\bibitem[Islam et~al\mbox{.}(2023)]%
        {Hindle:JavaTestEvolution:MSR:2023}
\bibfield{author}{\bibinfo{person}{Anisha Islam}, \bibinfo{person}{Nipuni~Tharushika Hewage}, \bibinfo{person}{Abdul Ali~Bangash}, {and} \bibinfo{person}{Abram Hindle}.} \bibinfo{year}{2023}\natexlab{}.
\newblock \showarticletitle{Evolution of the Practice of Software Testing in Java Projects}. In \bibinfo{booktitle}{\emph{2023 IEEE/ACM 20th International Conference on Mining Software Repositories (MSR)}}. \bibinfo{pages}{367--371}.
\newblock
\urldef\tempurl%
\url{https://doi.org/10.1109/MSR59073.2023.00057}
\showDOI{\tempurl}


\bibitem[Just et~al\mbox{.}(2014)]%
        {Just2014}
\bibfield{author}{\bibinfo{person}{Ren\'{e} Just}, \bibinfo{person}{Darioush Jalali}, {and} \bibinfo{person}{Michael~D. Ernst}.} \bibinfo{year}{2014}\natexlab{}.
\newblock \showarticletitle{Defects4J: a database of existing faults to enable controlled testing studies for Java programs}. In \bibinfo{booktitle}{\emph{Proceedings of the 2014 International Symposium on Software Testing and Analysis}} (San Jose, CA, USA) \emph{(\bibinfo{series}{ISSTA 2014})}. \bibinfo{publisher}{Association for Computing Machinery}, \bibinfo{address}{New York, NY, USA}, \bibinfo{pages}{437–440}.
\newblock
\showISBNx{9781450326452}
\urldef\tempurl%
\url{https://doi.org/10.1145/2610384.2628055}
\showDOI{\tempurl}


\bibitem[Kang et~al\mbox{.}(2023)]%
        {Kang:FewShotTesters:ICSE:2023}
\bibfield{author}{\bibinfo{person}{Sungmin Kang}, \bibinfo{person}{Juyeon Yoon}, {and} \bibinfo{person}{Shin Yoo}.} \bibinfo{year}{2023}\natexlab{}.
\newblock \showarticletitle{Large Language Models are Few-Shot Testers: Exploring LLM-Based General Bug Reproduction}. In \bibinfo{booktitle}{\emph{Proceedings of the 45th International Conference on Software Engineering}} (Melbourne, Victoria, Australia) \emph{(\bibinfo{series}{ICSE '23})}. \bibinfo{publisher}{IEEE Press}, \bibinfo{pages}{2312–2323}.
\newblock
\showISBNx{9781665457019}
\urldef\tempurl%
\url{https://doi.org/10.1109/ICSE48619.2023.00194}
\showDOI{\tempurl}


\bibitem[Kochhar et~al\mbox{.}(2015)]%
        {Kochhar:TestSuiteEffectiveness:SANER:2015}
\bibfield{author}{\bibinfo{person}{Pavneet~Singh Kochhar}, \bibinfo{person}{Ferdian Thung}, {and} \bibinfo{person}{David Lo}.} \bibinfo{year}{2015}\natexlab{}.
\newblock \showarticletitle{Code coverage and test suite effectiveness: Empirical study with real bugs in large systems}. In \bibinfo{booktitle}{\emph{2015 IEEE 22nd International Conference on Software Analysis, Evolution, and Reengineering (SANER)}}. \bibinfo{pages}{560--564}.
\newblock
\urldef\tempurl%
\url{https://doi.org/10.1109/SANER.2015.7081877}
\showDOI{\tempurl}


\bibitem[Lei et~al\mbox{.}(2024)]%
        {Lei2024}
\bibfield{author}{\bibinfo{person}{Zhanyao Lei}, \bibinfo{person}{Yixiong Chen}, \bibinfo{person}{Mingyuan Xia}, {and} \bibinfo{person}{Zhengwei Qi}.} \bibinfo{year}{2024}\natexlab{}.
\newblock \showarticletitle{Foliage: Nourishing Evolving Software by Characterizing and Clustering Field Bugs}. In \bibinfo{booktitle}{\emph{Proceedings of the 33rd ACM SIGSOFT International Symposium on Software Testing and Analysis}} (Vienna, Austria) \emph{(\bibinfo{series}{ISSTA 2024})}. \bibinfo{publisher}{Association for Computing Machinery}, \bibinfo{address}{New York, NY, USA}, \bibinfo{pages}{1325–1337}.
\newblock
\showISBNx{9798400706127}
\urldef\tempurl%
\url{https://doi.org/10.1145/3650212.3680363}
\showDOI{\tempurl}


\bibitem[Lewis et~al\mbox{.}(2020)]%
        {Lewis:RetrievalAugmentedGeneration:NIPS:2020}
\bibfield{author}{\bibinfo{person}{Patrick Lewis}, \bibinfo{person}{Ethan Perez}, \bibinfo{person}{Aleksandra Piktus}, \bibinfo{person}{Fabio Petroni}, \bibinfo{person}{Vladimir Karpukhin}, \bibinfo{person}{Naman Goyal}, \bibinfo{person}{Heinrich K\"{u}ttler}, \bibinfo{person}{Mike Lewis}, \bibinfo{person}{Wen-tau Yih}, \bibinfo{person}{Tim Rockt\"{a}schel}, \bibinfo{person}{Sebastian Riedel}, {and} \bibinfo{person}{Douwe Kiela}.} \bibinfo{year}{2020}\natexlab{}.
\newblock \showarticletitle{Retrieval-augmented generation for knowledge-intensive NLP tasks}. In \bibinfo{booktitle}{\emph{Proceedings of the 34th International Conference on Neural Information Processing Systems}} (Vancouver, BC, Canada) \emph{(\bibinfo{series}{NIPS '20})}. \bibinfo{publisher}{Curran Associates Inc.}, \bibinfo{address}{Red Hook, NY, USA}, Article \bibinfo{articleno}{793}, \bibinfo{numpages}{16}~pages.
\newblock
\showISBNx{9781713829546}


\bibitem[Li et~al\mbox{.}(2023)]%
        {Li:DJXPerf:CGO:2023}
\bibfield{author}{\bibinfo{person}{Bolun Li}, \bibinfo{person}{Pengfei Su}, \bibinfo{person}{Milind Chabbi}, \bibinfo{person}{Shuyin Jiao}, {and} \bibinfo{person}{Xu Liu}.} \bibinfo{year}{2023}\natexlab{}.
\newblock \showarticletitle{DJXPerf: Identifying Memory Inefficiencies via Object-Centric Profiling for Java}. In \bibinfo{booktitle}{\emph{Proceedings of the 21st ACM/IEEE International Symposium on Code Generation and Optimization}} (Montr\'{e}al, QC, Canada) \emph{(\bibinfo{series}{CGO '23})}. \bibinfo{publisher}{Association for Computing Machinery}, \bibinfo{address}{New York, NY, USA}, \bibinfo{pages}{81–94}.
\newblock
\showISBNx{9798400701016}
\urldef\tempurl%
\url{https://doi.org/10.1145/3579990.3580010}
\showDOI{\tempurl}


\bibitem[Liu et~al\mbox{.}(2023)]%
        {Liu2023}
\bibfield{author}{\bibinfo{person}{Yu Liu}, \bibinfo{person}{Jiyang Zhang}, \bibinfo{person}{Pengyu Nie}, \bibinfo{person}{Milos Gligoric}, {and} \bibinfo{person}{Owolabi Legunsen}.} \bibinfo{year}{2023}\natexlab{}.
\newblock \showarticletitle{More Precise Regression Test Selection via Reasoning about Semantics-Modifying Changes}. In \bibinfo{booktitle}{\emph{Proceedings of the 32nd ACM SIGSOFT International Symposium on Software Testing and Analysis}} (Seattle, WA, USA) \emph{(\bibinfo{series}{ISSTA 2023})}. \bibinfo{publisher}{Association for Computing Machinery}, \bibinfo{address}{New York, NY, USA}, \bibinfo{pages}{664–676}.
\newblock
\showISBNx{9798400702211}
\urldef\tempurl%
\url{https://doi.org/10.1145/3597926.3598086}
\showDOI{\tempurl}


\bibitem[Liu et~al\mbox{.}(2024)]%
        {Liu:TestProgramMutation:ASE:2024}
\bibfield{author}{\bibinfo{person}{Yujie Liu}, \bibinfo{person}{Mingxuan Zhu}, \bibinfo{person}{Jinhao Dong}, \bibinfo{person}{Junzhe Yu}, {and} \bibinfo{person}{Dan Hao}.} \bibinfo{year}{2024}\natexlab{}.
\newblock \showarticletitle{Compiler Bug Isolation via Enhanced Test Program Mutation}. In \bibinfo{booktitle}{\emph{Proceedings of the 39th IEEE/ACM International Conference on Automated Software Engineering}} (Sacramento, CA, USA) \emph{(\bibinfo{series}{ASE '24})}. \bibinfo{publisher}{Association for Computing Machinery}, \bibinfo{address}{New York, NY, USA}, \bibinfo{pages}{819–830}.
\newblock
\showISBNx{9798400712487}
\urldef\tempurl%
\url{https://doi.org/10.1145/3691620.3695074}
\showDOI{\tempurl}


\bibitem[Lu et~al\mbox{.}(2024)]%
        {Lu:DiaVio:ISSTA:2024}
\bibfield{author}{\bibinfo{person}{You Lu}, \bibinfo{person}{Yifan Tian}, \bibinfo{person}{Yuyang Bi}, \bibinfo{person}{Bihuan Chen}, {and} \bibinfo{person}{Xin Peng}.} \bibinfo{year}{2024}\natexlab{}.
\newblock \showarticletitle{DiaVio: LLM-Empowered Diagnosis of Safety Violations in ADS Simulation Testing}. In \bibinfo{booktitle}{\emph{Proceedings of the 33rd ACM SIGSOFT International Symposium on Software Testing and Analysis}} (Vienna, Austria) \emph{(\bibinfo{series}{ISSTA 2024})}. \bibinfo{publisher}{Association for Computing Machinery}, \bibinfo{address}{New York, NY, USA}, \bibinfo{pages}{376–388}.
\newblock
\showISBNx{9798400706127}
\urldef\tempurl%
\url{https://doi.org/10.1145/3650212.3652135}
\showDOI{\tempurl}


\bibitem[Ma et~al\mbox{.}(2021)]%
        {Ma:TestSelection:TOSEM:2021}
\bibfield{author}{\bibinfo{person}{Wei Ma}, \bibinfo{person}{Mike Papadakis}, \bibinfo{person}{Anestis Tsakmalis}, \bibinfo{person}{Maxime Cordy}, {and} \bibinfo{person}{Yves~Le Traon}.} \bibinfo{year}{2021}\natexlab{}.
\newblock \showarticletitle{Test Selection for Deep Learning Systems}.
\newblock \bibinfo{journal}{\emph{ACM Trans. Softw. Eng. Methodol.}} \bibinfo{volume}{30}, \bibinfo{number}{2}, Article \bibinfo{articleno}{13} (\bibinfo{date}{jan} \bibinfo{year}{2021}), \bibinfo{numpages}{22}~pages.
\newblock
\showISSN{1049-331X}
\urldef\tempurl%
\url{https://doi.org/10.1145/3417330}
\showDOI{\tempurl}


\bibitem[McKenzie et~al\mbox{.}(2023)]%
        {mckenzie2023inverse}
\bibfield{author}{\bibinfo{person}{Ian~R. McKenzie}, \bibinfo{person}{Alexander Lyzhov}, \bibinfo{person}{Michael~Martin Pieler}, \bibinfo{person}{Alicia Parrish}, \bibinfo{person}{Aaron Mueller}, \bibinfo{person}{Ameya Prabhu}, \bibinfo{person}{Euan McLean}, \bibinfo{person}{Xudong Shen}, \bibinfo{person}{Joe Cavanagh}, \bibinfo{person}{Andrew~George Gritsevskiy}, \bibinfo{person}{Derik Kauffman}, \bibinfo{person}{Aaron~T. Kirtland}, \bibinfo{person}{Zhengping Zhou}, \bibinfo{person}{Yuhui Zhang}, \bibinfo{person}{Sicong Huang}, \bibinfo{person}{Daniel Wurgaft}, \bibinfo{person}{Max Weiss}, \bibinfo{person}{Alexis Ross}, \bibinfo{person}{Gabriel Recchia}, \bibinfo{person}{Alisa Liu}, \bibinfo{person}{Jiacheng Liu}, \bibinfo{person}{Tom Tseng}, \bibinfo{person}{Tomasz Korbak}, \bibinfo{person}{Najoung Kim}, \bibinfo{person}{Samuel~R. Bowman}, {and} \bibinfo{person}{Ethan Perez}.} \bibinfo{year}{2023}\natexlab{}.
\newblock \showarticletitle{Inverse Scaling: When Bigger Isn't Better}.
\newblock \bibinfo{journal}{\emph{Transactions on Machine Learning Research}} (\bibinfo{year}{2023}).
\newblock
\showISSN{2835-8856}
\urldef\tempurl%
\url{https://openreview.net/forum?id=DwgRm72GQF}
\showURL{%
\tempurl}
\newblock
\shownote{Featured Certification}.


\bibitem[Memon et~al\mbox{.}(2017)]%
        {Memon2017}
\bibfield{author}{\bibinfo{person}{Atif Memon}, \bibinfo{person}{Zebao Gao}, \bibinfo{person}{Bao Nguyen}, \bibinfo{person}{Sanjeev Dhanda}, \bibinfo{person}{Eric Nickell}, \bibinfo{person}{Rob Siemborski}, {and} \bibinfo{person}{John Micco}.} \bibinfo{year}{2017}\natexlab{}.
\newblock \showarticletitle{Taming Google-scale continuous testing}. In \bibinfo{booktitle}{\emph{2017 IEEE/ACM 39th International Conference on Software Engineering: Software Engineering in Practice Track (ICSE-SEIP)}}. \bibinfo{pages}{233--242}.
\newblock
\urldef\tempurl%
\url{https://doi.org/10.1109/ICSE-SEIP.2017.16}
\showDOI{\tempurl}


\bibitem[Morán et~al\mbox{.}(2017)]%
        {moran2017}
\bibfield{author}{\bibinfo{person}{Jesús Morán}, \bibinfo{person}{Antonia Bertolino}, \bibinfo{person}{Claudio de~la Riva}, {and} \bibinfo{person}{Javier Tuya}.} \bibinfo{year}{2017}\natexlab{}.
\newblock \showarticletitle{Towards Ex Vivo Testing of MapReduce Applications}. In \bibinfo{booktitle}{\emph{2017 IEEE International Conference on Software Quality, Reliability and Security (QRS)}}. \bibinfo{pages}{73--80}.
\newblock
\urldef\tempurl%
\url{https://doi.org/10.1109/QRS.2017.17}
\showDOI{\tempurl}


\bibitem[Murphy et~al\mbox{.}(2009)]%
        {murphy2009}
\bibfield{author}{\bibinfo{person}{Christian Murphy}, \bibinfo{person}{Gail Kaiser}, \bibinfo{person}{Ian Vo}, {and} \bibinfo{person}{Matt Chu}.} \bibinfo{year}{2009}\natexlab{}.
\newblock \showarticletitle{Quality Assurance of Software Applications Using the In Vivo Testing Approach}. In \bibinfo{booktitle}{\emph{2009 International Conference on Software Testing Verification and Validation}}. \bibinfo{pages}{111--120}.
\newblock
\urldef\tempurl%
\url{https://doi.org/10.1109/ICST.2009.18}
\showDOI{\tempurl}


\bibitem[Nong et~al\mbox{.}(2025)]%
        {Nong:APPATCH:USENIX:2025}
\bibfield{author}{\bibinfo{person}{Yu Nong}, \bibinfo{person}{Haoran Yang}, \bibinfo{person}{Long Cheng}, \bibinfo{person}{Honxin Hu}, {and} \bibinfo{person}{Haipeng Cai}.} \bibinfo{year}{2025}\natexlab{}.
\newblock \showarticletitle{APPATCH: Automated Adaptive Prompting Large Language Models for Real-World Software Vulnerability Patching}. In \bibinfo{booktitle}{\emph{34th USENIX Security Symposium (USENIX Security 25)}}. \bibinfo{pages}{1--20}.
\newblock


\bibitem[Opitz and Burst(2019)]%
        {opitz2019macro}
\bibfield{author}{\bibinfo{person}{Juri Opitz} {and} \bibinfo{person}{Sebastian Burst}.} \bibinfo{year}{2019}\natexlab{}.
\newblock \showarticletitle{Macro f1 and macro f1}.
\newblock \bibinfo{journal}{\emph{arXiv preprint arXiv:1911.03347}} (\bibinfo{year}{2019}).
\newblock


\bibitem[Ouyang et~al\mbox{.}(2025)]%
        {Ouyang:ChatGPTCodeGeneration:TOSEM:2025}
\bibfield{author}{\bibinfo{person}{Shuyin Ouyang}, \bibinfo{person}{Jie~M. Zhang}, \bibinfo{person}{Mark Harman}, {and} \bibinfo{person}{Meng Wang}.} \bibinfo{year}{2025}\natexlab{}.
\newblock \showarticletitle{An Empirical Study of the Non-Determinism of ChatGPT in Code Generation}.
\newblock \bibinfo{journal}{\emph{ACM Trans. Softw. Eng. Methodol.}} \bibinfo{volume}{34}, \bibinfo{number}{2}, Article \bibinfo{articleno}{42} (\bibinfo{date}{Jan.} \bibinfo{year}{2025}), \bibinfo{numpages}{28}~pages.
\newblock
\showISSN{1049-331X}
\urldef\tempurl%
\url{https://doi.org/10.1145/3697010}
\showDOI{\tempurl}


\bibitem[Pan et~al\mbox{.}(2022)]%
        {Pan:TestCaseSelection:EMSE:2022}
\bibfield{author}{\bibinfo{person}{Rongqi Pan}, \bibinfo{person}{Mojtaba Bagherzadeh}, \bibinfo{person}{Taher~A. Ghaleb}, {and} \bibinfo{person}{Lionel Briand}.} \bibinfo{year}{2022}\natexlab{}.
\newblock \showarticletitle{Test case selection and prioritization using machine learning: a systematic literature review}.
\newblock \bibinfo{journal}{\emph{Empirical Softw. Engg.}} \bibinfo{volume}{27}, \bibinfo{number}{2} (\bibinfo{date}{mar} \bibinfo{year}{2022}), \bibinfo{numpages}{43}~pages.
\newblock
\showISSN{1382-3256}
\urldef\tempurl%
\url{https://doi.org/10.1007/s10664-021-10066-6}
\showDOI{\tempurl}


\bibitem[Pei et~al\mbox{.}(2023)]%
        {Pei:LLMProgramInvariants:ICML:2023}
\bibfield{author}{\bibinfo{person}{Kexin Pei}, \bibinfo{person}{David Bieber}, \bibinfo{person}{Kensen Shi}, \bibinfo{person}{Charles Sutton}, {and} \bibinfo{person}{Pengcheng Yin}.} \bibinfo{year}{2023}\natexlab{}.
\newblock \showarticletitle{Can large language models reason about program invariants?}. In \bibinfo{booktitle}{\emph{Proceedings of the 40th International Conference on Machine Learning}} (Honolulu, Hawaii, USA) \emph{(\bibinfo{series}{ICML'23})}. \bibinfo{publisher}{JMLR.org}, Article \bibinfo{articleno}{1144}, \bibinfo{numpages}{25}~pages.
\newblock


\bibitem[Pezz{\`e} and Young(2008)]%
        {pezze2008}
\bibfield{author}{\bibinfo{person}{Mauro Pezz{\`e}} {and} \bibinfo{person}{Michal Young}.} \bibinfo{year}{2008}\natexlab{}.
\newblock \bibinfo{booktitle}{\emph{Software testing and analysis: process, principles, and techniques}}.
\newblock \bibinfo{publisher}{John Wiley \& Sons}.
\newblock


\bibitem[Pham et~al\mbox{.}(2023)]%
        {pham2022}
\bibfield{author}{\bibinfo{person}{Phuoc Pham}, \bibinfo{person}{Vu Nguyen}, {and} \bibinfo{person}{Tien Nguyen}.} \bibinfo{year}{2023}\natexlab{}.
\newblock \showarticletitle{A Review of AI-augmented End-to-End Test Automation Tools}. In \bibinfo{booktitle}{\emph{Proceedings of the 37th IEEE/ACM International Conference on Automated Software Engineering}} (Rochester, MI, USA) \emph{(\bibinfo{series}{ASE '22})}. \bibinfo{publisher}{Association for Computing Machinery}, \bibinfo{address}{New York, NY, USA}, Article \bibinfo{articleno}{214}, \bibinfo{numpages}{4}~pages.
\newblock
\showISBNx{9781450394758}
\urldef\tempurl%
\url{https://doi.org/10.1145/3551349.3563240}
\showDOI{\tempurl}


\bibitem[Pinto et~al\mbox{.}(2012)]%
        {Pinto:TestSuiteEvolution:FSE:2012}
\bibfield{author}{\bibinfo{person}{Leandro~Sales Pinto}, \bibinfo{person}{Saurabh Sinha}, {and} \bibinfo{person}{Alessandro Orso}.} \bibinfo{year}{2012}\natexlab{}.
\newblock \showarticletitle{Understanding myths and realities of test-suite evolution}. In \bibinfo{booktitle}{\emph{Proceedings of the ACM SIGSOFT 20th International Symposium on the Foundations of Software Engineering}} (Cary, North Carolina) \emph{(\bibinfo{series}{FSE '12})}. \bibinfo{publisher}{Association for Computing Machinery}, \bibinfo{address}{New York, NY, USA}, Article \bibinfo{articleno}{33}, \bibinfo{numpages}{11}~pages.
\newblock
\showISBNx{9781450316149}
\urldef\tempurl%
\url{https://doi.org/10.1145/2393596.2393634}
\showDOI{\tempurl}


\bibitem[Porter et~al\mbox{.}(2007)]%
        {Porter:Skoll:TSE:2007}
\bibfield{author}{\bibinfo{person}{Adam Porter}, \bibinfo{person}{Cemal Yilmaz}, \bibinfo{person}{Atif~M. Memon}, \bibinfo{person}{Douglas~C. Schmidt}, {and} \bibinfo{person}{Bala Natarajan}.} \bibinfo{year}{2007}\natexlab{}.
\newblock \showarticletitle{Skoll: A Process and Infrastructure for Distributed Continuous Quality Assurance}.
\newblock \bibinfo{journal}{\emph{IEEE Transactions on Software Engineering}} \bibinfo{volume}{33}, \bibinfo{number}{8} (\bibinfo{year}{2007}), \bibinfo{pages}{510--525}.
\newblock
\urldef\tempurl%
\url{https://doi.org/10.1109/TSE.2007.70719}
\showDOI{\tempurl}


\bibitem[Richardson and Clarke(1981)]%
        {Richardson:partion:ICSE:1981}
\bibfield{author}{\bibinfo{person}{Debra~J. Richardson} {and} \bibinfo{person}{Lori~A. Clarke}.} \bibinfo{year}{1981}\natexlab{}.
\newblock \showarticletitle{A partition analysis method to increase program reliability}. In \bibinfo{booktitle}{\emph{Proceedings of the 5th International Conference on Software Engineering}} (San Diego, California, USA) \emph{(\bibinfo{series}{ICSE '81})}. \bibinfo{publisher}{IEEE Press}, \bibinfo{pages}{244–253}.
\newblock
\showISBNx{0897911466}


\bibitem[Santelices et~al\mbox{.}(2008)]%
        {Santelices:TestSuiteAugmentation:ASE:2008}
\bibfield{author}{\bibinfo{person}{Raul Santelices}, \bibinfo{person}{Pavan~Kumar Chittimalli}, \bibinfo{person}{Taweesup Apiwattanapong}, \bibinfo{person}{Alessandro Orso}, {and} \bibinfo{person}{Mary~Jean Harrold}.} \bibinfo{year}{2008}\natexlab{}.
\newblock \showarticletitle{Test-Suite Augmentation for Evolving Software}. In \bibinfo{booktitle}{\emph{2008 23rd IEEE/ACM International Conference on Automated Software Engineering}}. \bibinfo{publisher}{IEEE Computer Society}, \bibinfo{address}{USA}, \bibinfo{pages}{218--227}.
\newblock
\showISBNx{9781424421879}
\urldef\tempurl%
\url{https://doi.org/10.1109/ASE.2008.32}
\showDOI{\tempurl}


\bibitem[Smith et~al\mbox{.}(2023)]%
        {javaparser}
\bibfield{author}{\bibinfo{person}{Nicholas Smith}, \bibinfo{person}{Danny van Bruggen}, {and} \bibinfo{person}{Federico Tomassetti}.} \bibinfo{year}{2023}\natexlab{}.
\newblock \bibinfo{title}{JavaParser: Visited}.
\newblock
\newblock
\urldef\tempurl%
\url{https://leanpub.com/javaparservisited}
\showURL{%
\tempurl}
\newblock
\shownote{Last accessed 22 October 2024}.


\bibitem[Su et~al\mbox{.}(2024)]%
        {Su:ExploratoryTestingLLM:ICSE:2024}
\bibfield{author}{\bibinfo{person}{Yanqi Su}, \bibinfo{person}{Dianshu Liao}, \bibinfo{person}{Zhenchang Xing}, \bibinfo{person}{Qing Huang}, \bibinfo{person}{Mulong Xie}, \bibinfo{person}{Qinghua Lu}, {and} \bibinfo{person}{Xiwei Xu}.} \bibinfo{year}{2024}\natexlab{}.
\newblock \showarticletitle{Enhancing Exploratory Testing by Large Language Model and Knowledge Graph}. In \bibinfo{booktitle}{\emph{Proceedings of the IEEE/ACM 46th International Conference on Software Engineering}} (Lisbon, Portugal) \emph{(\bibinfo{series}{ICSE '24})}. \bibinfo{publisher}{Association for Computing Machinery}, \bibinfo{address}{New York, NY, USA}, Article \bibinfo{articleno}{98}, \bibinfo{numpages}{12}~pages.
\newblock
\showISBNx{9798400702174}
\urldef\tempurl%
\url{https://doi.org/10.1145/3597503.3639157}
\showDOI{\tempurl}


\bibitem[Wang et~al\mbox{.}(2024)]%
        {wang2024}
\bibfield{author}{\bibinfo{person}{Junjie Wang}, \bibinfo{person}{Yuchao Huang}, \bibinfo{person}{Chunyang Chen}, \bibinfo{person}{Zhe Liu}, \bibinfo{person}{Song Wang}, {and} \bibinfo{person}{Qing Wang}.} \bibinfo{year}{2024}\natexlab{}.
\newblock \showarticletitle{Software Testing With Large Language Models: Survey, Landscape, and Vision}.
\newblock \bibinfo{journal}{\emph{IEEE Trans. Softw. Eng.}} \bibinfo{volume}{50}, \bibinfo{number}{4} (\bibinfo{date}{April} \bibinfo{year}{2024}), \bibinfo{pages}{911–936}.
\newblock
\showISSN{0098-5589}
\urldef\tempurl%
\url{https://doi.org/10.1109/TSE.2024.3368208}
\showDOI{\tempurl}


\bibitem[Wang et~al\mbox{.}(2025)]%
        {Wang:TESTEVAL:2025}
\bibfield{author}{\bibinfo{person}{Wenhan Wang}, \bibinfo{person}{Chenyuan Yang}, \bibinfo{person}{Zhijie Wang}, \bibinfo{person}{Yuheng Huang}, \bibinfo{person}{Zhaoyang Chu}, \bibinfo{person}{Da Song}, \bibinfo{person}{Lingming Zhang}, \bibinfo{person}{An~Ran Chen}, {and} \bibinfo{person}{Lei Ma}.} \bibinfo{year}{2025}\natexlab{}.
\newblock \bibinfo{title}{TESTEVAL: Benchmarking Large Language Models for Test Case Generation}.
\newblock
\newblock
\showeprint[arxiv]{2406.04531}~[cs.SE]
\urldef\tempurl%
\url{https://arxiv.org/abs/2406.04531}
\showURL{%
\tempurl}


\bibitem[Wang et~al\mbox{.}(2020)]%
        {Wang:SurveyFewshotLearning:ACMCS:2020}
\bibfield{author}{\bibinfo{person}{Yaqing Wang}, \bibinfo{person}{Quanming Yao}, \bibinfo{person}{James~T. Kwok}, {and} \bibinfo{person}{Lionel~M. Ni}.} \bibinfo{year}{2020}\natexlab{}.
\newblock \showarticletitle{Generalizing from a Few Examples: A Survey on Few-shot Learning}.
\newblock \bibinfo{journal}{\emph{ACM Comput. Surv.}} \bibinfo{volume}{53}, \bibinfo{number}{3}, Article \bibinfo{articleno}{63} (\bibinfo{date}{June} \bibinfo{year}{2020}), \bibinfo{numpages}{34}~pages.
\newblock
\showISSN{0360-0300}
\urldef\tempurl%
\url{https://doi.org/10.1145/3386252}
\showDOI{\tempurl}


\bibitem[Weyssow et~al\mbox{.}(2025)]%
        {Weyssow:FineTuningCodeGenerationLargeLanguageModels:TOSEM:2025}
\bibfield{author}{\bibinfo{person}{Martin Weyssow}, \bibinfo{person}{Xin Zhou}, \bibinfo{person}{Kisub Kim}, \bibinfo{person}{David Lo}, {and} \bibinfo{person}{Houari Sahraoui}.} \bibinfo{year}{2025}\natexlab{}.
\newblock \showarticletitle{Exploring Parameter-Efficient Fine-Tuning Techniques for Code Generation with Large Language Models}.
\newblock \bibinfo{journal}{\emph{ACM Trans. Softw. Eng. Methodol.}} (\bibinfo{date}{Jan.} \bibinfo{year}{2025}).
\newblock
\showISSN{1049-331X}
\urldef\tempurl%
\url{https://doi.org/10.1145/3714461}
\showDOI{\tempurl}
\newblock
\shownote{Just Accepted}.


\bibitem[Winkler et~al\mbox{.}(2022)]%
        {winkler2022we}
\bibfield{author}{\bibinfo{person}{Dietmar Winkler}, \bibinfo{person}{Pirmin Urbanke}, {and} \bibinfo{person}{Rudolf Ramler}.} \bibinfo{year}{2022}\natexlab{}.
\newblock \showarticletitle{What do we know about readability of test code?-a systematic mapping study}. In \bibinfo{booktitle}{\emph{2022 IEEE International Conference on Software Analysis, Evolution and Reengineering (SANER)}}. IEEE, \bibinfo{pages}{1167--1174}.
\newblock


\bibitem[Xu et~al\mbox{.}(2010)]%
        {Xu:DirectedTestSuiteAugmentationTechniquesAndTradeoffs:FSE:2010}
\bibfield{author}{\bibinfo{person}{Zhihong Xu}, \bibinfo{person}{Yunho Kim}, \bibinfo{person}{Moonzoo Kim}, \bibinfo{person}{Gregg Rothermel}, {and} \bibinfo{person}{Myra~B. Cohen}.} \bibinfo{year}{2010}\natexlab{}.
\newblock \showarticletitle{Directed test suite augmentation: techniques and tradeoffs}. In \bibinfo{booktitle}{\emph{Proceedings of the Eighteenth ACM SIGSOFT International Symposium on Foundations of Software Engineering}} (Santa Fe, New Mexico, USA) \emph{(\bibinfo{series}{FSE '10})}. \bibinfo{publisher}{Association for Computing Machinery}, \bibinfo{address}{New York, NY, USA}, \bibinfo{pages}{257–266}.
\newblock
\showISBNx{9781605587912}
\urldef\tempurl%
\url{https://doi.org/10.1145/1882291.1882330}
\showDOI{\tempurl}


\bibitem[Xue et~al\mbox{.}(2025)]%
        {xue2025}
\bibfield{author}{\bibinfo{person}{Pengyu Xue}, \bibinfo{person}{Linhao Wu}, \bibinfo{person}{Zhen Yang}, \bibinfo{person}{Chengyi Wang}, \bibinfo{person}{Xiang Li}, \bibinfo{person}{Yuxiang Zhang}, \bibinfo{person}{Jia Li}, \bibinfo{person}{Ruikai Jin}, \bibinfo{person}{Yifei Pei}, \bibinfo{person}{Zhaoyan Shen}, {et~al\mbox{.}}} \bibinfo{year}{2025}\natexlab{}.
\newblock \showarticletitle{ClassEval-T: Evaluating Large Language Models in Class-Level Code Translation}.
\newblock \bibinfo{journal}{\emph{Proceedings of the ACM on Software Engineering}} \bibinfo{volume}{2}, \bibinfo{number}{ISSTA} (\bibinfo{year}{2025}), \bibinfo{pages}{1421--1444}.
\newblock


\bibitem[Yan et~al\mbox{.}(2023)]%
        {yan-etal-2023-codetransocean}
\bibfield{author}{\bibinfo{person}{Weixiang Yan}, \bibinfo{person}{Yuchen Tian}, \bibinfo{person}{Yunzhe Li}, \bibinfo{person}{Qian Chen}, {and} \bibinfo{person}{Wen Wang}.} \bibinfo{year}{2023}\natexlab{}.
\newblock \showarticletitle{{C}ode{T}rans{O}cean: A Comprehensive Multilingual Benchmark for Code Translation}. In \bibinfo{booktitle}{\emph{Findings of the Association for Computational Linguistics: EMNLP 2023}}, \bibfield{editor}{\bibinfo{person}{Houda Bouamor}, \bibinfo{person}{Juan Pino}, {and} \bibinfo{person}{Kalika Bali}} (Eds.). \bibinfo{publisher}{Association for Computational Linguistics}, \bibinfo{address}{Singapore}, \bibinfo{pages}{5067--5089}.
\newblock
\urldef\tempurl%
\url{https://doi.org/10.18653/v1/2023.findings-emnlp.337}
\showDOI{\tempurl}


\bibitem[Yang et~al\mbox{.}(2024b)]%
        {Yang:LLMUnitTestGeneration:ASE:2024}
\bibfield{author}{\bibinfo{person}{Lin Yang}, \bibinfo{person}{Chen Yang}, \bibinfo{person}{Shutao Gao}, \bibinfo{person}{Weijing Wang}, \bibinfo{person}{Bo Wang}, \bibinfo{person}{Qihao Zhu}, \bibinfo{person}{Xiao Chu}, \bibinfo{person}{Jianyi Zhou}, \bibinfo{person}{Guangtai Liang}, \bibinfo{person}{Qianxiang Wang}, {and} \bibinfo{person}{Junjie Chen}.} \bibinfo{year}{2024}\natexlab{b}.
\newblock \showarticletitle{On the Evaluation of Large Language Models in Unit Test Generation}. In \bibinfo{booktitle}{\emph{Proceedings of the 39th IEEE/ACM International Conference on Automated Software Engineering}} (Sacramento, CA, USA) \emph{(\bibinfo{series}{ASE '24})}. \bibinfo{publisher}{Association for Computing Machinery}, \bibinfo{address}{New York, NY, USA}, \bibinfo{pages}{1607–1619}.
\newblock
\showISBNx{9798400712487}
\urldef\tempurl%
\url{https://doi.org/10.1145/3691620.3695529}
\showDOI{\tempurl}


\bibitem[Yang et~al\mbox{.}(2024a)]%
        {Yang:SparseCoder:ESE:2024}
\bibfield{author}{\bibinfo{person}{Xueqi Yang}, \bibinfo{person}{Mariusz Jakubowski}, \bibinfo{person}{Li Kang}, \bibinfo{person}{Haojie Yu}, {and} \bibinfo{person}{Tim Menzies}.} \bibinfo{year}{2024}\natexlab{a}.
\newblock \showarticletitle{SparseCoder: Advancing source code analysis with sparse attention and learned token pruning}.
\newblock \bibinfo{journal}{\emph{Empirical Softw. Engg.}} \bibinfo{volume}{30}, \bibinfo{number}{1} (\bibinfo{date}{Dec.} \bibinfo{year}{2024}), \bibinfo{numpages}{30}~pages.
\newblock
\showISSN{1382-3256}
\urldef\tempurl%
\url{https://doi.org/10.1007/s10664-024-10558-1}
\showDOI{\tempurl}


\bibitem[Yoo and Harman(2012)]%
        {yoo2012regression}
\bibfield{author}{\bibinfo{person}{Shin Yoo} {and} \bibinfo{person}{Mark Harman}.} \bibinfo{year}{2012}\natexlab{}.
\newblock \showarticletitle{Regression testing minimization, selection and prioritization: a survey}.
\newblock \bibinfo{journal}{\emph{Software testing, verification and reliability}} \bibinfo{volume}{22}, \bibinfo{number}{2} (\bibinfo{year}{2012}), \bibinfo{pages}{67--120}.
\newblock


\bibitem[Zhang et~al\mbox{.}(2011)]%
        {Zhang:JunitTestSuiteReduction:ISSRE:2011}
\bibfield{author}{\bibinfo{person}{Lingming Zhang}, \bibinfo{person}{Darko Marinov}, \bibinfo{person}{Lu Zhang}, {and} \bibinfo{person}{Sarfraz Khurshid}.} \bibinfo{year}{2011}\natexlab{}.
\newblock \showarticletitle{An Empirical Study of JUnit Test-Suite Reduction}. In \bibinfo{booktitle}{\emph{Proceedings of the 2011 IEEE 22nd International Symposium on Software Reliability Engineering}} \emph{(\bibinfo{series}{ISSRE '11})}. \bibinfo{publisher}{IEEE Computer Society}, \bibinfo{address}{USA}, \bibinfo{pages}{170–179}.
\newblock
\showISBNx{9780769545684}
\urldef\tempurl%
\url{https://doi.org/10.1109/ISSRE.2011.26}
\showDOI{\tempurl}


\bibitem[Zhou et~al\mbox{.}(2016)]%
        {Zhou:MetamorphicTesting:TSE:2016}
\bibfield{author}{\bibinfo{person}{Zhi~Quan Zhou}, \bibinfo{person}{Shaowen Xiang}, {and} \bibinfo{person}{Tsong~Yueh Chen}.} \bibinfo{year}{2016}\natexlab{}.
\newblock \showarticletitle{Metamorphic Testing for Software Quality Assessment: A Study of Search Engines}.
\newblock \bibinfo{journal}{\emph{IEEE Transactions on Software Engineering}} \bibinfo{volume}{42}, \bibinfo{number}{3} (\bibinfo{year}{2016}), \bibinfo{pages}{264--284}.
\newblock
\urldef\tempurl%
\url{https://doi.org/10.1109/TSE.2015.2478001}
\showDOI{\tempurl}


\end{thebibliography}

\end{document}